# The Unsettled Number: Hubble's Tension

**Jorge L. Cervantes-Cota [1,*], Salvador Galindo-Uribarri [1,†] and George F. Smoot [2,3,4,5]**

[1] Department of Physics, National Institute for Nuclear Research, Km 36.5 Carretera Mexico-Toluca, Ocoyoacac C.P. 52750, State of Mexico, Mexico
[2] Donostia International Physics Center DIPC, Basque Country, San Sebastian, E-48080 Spain; gfsmoot@lbl.gov
[3] Institute for Advanced Study, Hong Kong University of Science and Technology, Clear Water Bay, Kowloon, Hong Kong 999077, China; Emeritus
[4] Université Sorbonne Paris Cité, Laboratoire APC-PCCP, Université Paris Diderot, 10 rue Alice Domon et Leonie Duquet, CEDEX 13, 75205 Paris, France; Emeritus
[5] Department of Physics and LBNL, University of California, MS Bldg 50-5505 LBNL, 1 Cyclotron Road, Berkeley, CA 94720, USA; Emeritus
* Correspondence: jorge.cervantes@inin.gob.mx
† Salvador Galindo-Uribarri passed away at the later stages of the manuscript preparation.

**Abstract:** One of main sources of uncertainty in modern cosmology is the present rate of the universe's expansion, $H_0$, called the Hubble constant. Once again, different observational techniques bring about different results, causing new "Hubble tension". In the present work, we review the historical roots of the Hubble constant from the beginning of the twentieth century, when modern cosmology originated, to the present. We develop the arguments that gave rise to the importance of measuring the expansion of the Universe and its discovery, and we describe the different pioneering works attempting to measure it. There has been a long dispute on this matter, even in the present epoch, which is marked by high-tech instrumentation and, therefore, in smaller uncertainties in the relevant parameters. It is, again, currently necessary to conduct a careful and critical revision of the different methods before one invokes new physics to solve the so-called Hubble tension.

---

## 1. Introduction

General relativity brought the idea of the expansion of space, not into something containing space, but of space itself. In the standard cosmological picture, the expansion rate of the universe is constantly changing as the cosmos evolves. The present-day expansion rate of the universe is given by a numerical value, the Hubble constant ($H_0$). This is a fixed number used as a unit of measurement to describe this expansion. The importance of knowing this value with good certainty lies in the fact that this knowledge provides a measurement scale of the present universe through the Hubble Sphere $r_{HS} = c/H_0$ and its time scale through Hubble time $t_H = 1/H_0 \sim 14$ Gyrs. It also serves to estimate the critical density of the universe, $\varrho_{critical} = 3\,H_0^2/8\,\pi\,G$, which, in general relativity, creates flat space geometry. Since $H_0$ is our least well-determined cosmological parameter, things are often expressed in terms of $h = H_0/(100$ km/s/Mpc), or $h^2$, in this case, so that the result can be scaled properly when accounting for final errors and results. Other cosmological parameters are intimately related to these quantities. Thus, the precise determination of $H_0$ could reveal missing pieces in our current understanding of physics. The present review provides a historical description of the origin, measurement, and present status of the Hubble constant.

During the past two decades, estimates of the constant have repeatedly been made using two different approaches. The first involved measuring features in the relative recent universe. The second approach used light left over from shortly after the big bang, the cosmic microwave background (CMB). These two routes are known, respectively, as the “late” and “early” routes to the Hubble constant, as the first involved measuring aspects of the late evolution stages of the universe, while the second focuses on the very early phase.

In recent years, a discrepancy regarding the estimation of $H_0$ has emerged, involving measurements from the two routes of its assessment. In defiance of all expectations, estimates of $H_0$ from early route assessments do not agree with those of the late route. This discrepancy is known as the Hubble tension. This tension might be nothing more than a measurement error. However, it has stubbornly prevailed in spite of the fact that all the recent measurements have increased their precision.

At present there is debate on whether experimental glitches in either set of estimates cause the discrepancy, but no one is sure what those glitches would be. However, there have been opinions that the Hubble tension points to something missing from our understanding of the cosmos that might require an adjustment of the standard cosmological model, also known as “ΛCDM”: cold dark matter plus constant dark energy Λ.

The scientific importance of resolving the tension has aroused the interest of scientists who are outsiders to the fields of astronomy and cosmology. This work is aimed at those who have interest in the Hubble tension, but work in fields of physical sciences other than those involved in the problem. It is hoped that this work will also appeal to field insiders, in particular, young postdoc researchers who are well-acquainted with the problem but would like to learn some historical details that are frequently missed or not mentioned in the current specialized literature. This present review might serve these individuals as a reference.

To achieve the purposes of this work, we have organized its narrative as follows:

We begin in Section 2, giving an account a debate held on April 1920 at the Smithsonian Museum of Natural History between Harlow Shapley and Heber Curtis on the size of the universe. Shapley maintained that the Milky Way was the entirety of the universe, while Curtis affirmed that the Milky Way was only one of multiple galaxies.

At the root of the debate was the problem of determining the distances at which celestial objects are situated. Thus, in Section 3, our story goes back to 1908, a few years before the debate, when Henrietta Leavitt discovered the relation between the luminosity and the period of Cepheid variables. Leavitt’s discovery provided astronomers with the first “standard candle” with which to measure the distances to faraway galaxies.

Section 4 centers its attention on Vesto Melvin Slipher. He was, in 1912, the first to discover that distant galaxies are redshifted, thus providing the first empirical basis for the expansion of the universe. Later, in the 1920s, Edwin Hubble showed that Andromeda was far outside the Milky Way by measuring Cepheid variable stars, proving that Curtis was correct. After Hubble found that the spiral nebulae were actually galaxies like ours, Slipher’s findings attracted the attention of some astronomers. On the one hand, most nebulae (now known to be galaxies) were rapidly moving away from Earth; on the other, a few were approaching it. Why did this imbalance exist? What was causing it?

Section 5 deals with De Sitter’s solution, found in 1917, to Einstein’s field equations for an empty universe. The solution implied an exponentially expanding universe. If De Sitter’s model were true, the redshifts observed by Slipher were indeed reflecting the predicted expansion of the universe. This suggested that a relation between redshifts and the distance of those light-emitting galaxies should exist.

Section 6 describes what Hubble found in 1929. What he found was a simple linear relation between the distances to galaxies and their recession speeds. He and his colleague Humason found the value of the constant of the linear relation to be about 500 km/s/Mpc. This was the first determination of what we know as the Hubble constant.

The next sections, Sections 7–10, provide an account of the emergence of cosmological relativity and the contributions of De Sitter, Friedmann, and Lemaître to information regarding the expanding universe. In 1926, Lemaître calculated the value $H_0$ to be about 625 km/s/Mpc.

Then, in Section 11, we come across “The earliest $H_0$-Tension”. This was a problem involving the age of the Earth based on early-20th radioactive dating as compared to the age of the universe, inferred from the value of the Hubble constant. It turned out that the Earth was found to be older than the Universe. Today, we know that the problem had its origin in the then-accepted value for $H_0$, which was far from today’s estimates.

The problem of correctly determining the value of the Hubble constant depends on the accurate assessment of distances to receding objects. Distances were once measured (and are still measured) by using Cepheids, which are high-luminosity variable stars. This method requires good calibration of the period P of the pulsating variable star to its maximum luminosity L (a P-L calibration). This is discussed in Section 12. Due to an odd coincidence, the P-L calibration used in the 1920s and 1930s was incorrect, consequently producing erroneous values for the Hubble constant.

A better calibration was found independently in the 1940s by Baade and Thackeray. This is presented in Section 13. As result, new attempts to measure the Hubble constant were triggered (Section 14). New values for the Hubble constant appeared, and during the period of 1940–1955, and the value of the constant reduced dramatically. A 1955 measurement by Humason, Mayall, and Sandage produced a value of 180 km/s/Mpc. With a stroke, this erased the issue of the supposed young age of the universe—it was older.

Section 15 mentions the need to measure $H_0$ at very large distances, since gravitational interactions among our neighboring galaxies may be causing some of them to move much more quickly or slowly than the rest of the universe.

This need led to the development of methods other than the use of Cepheids to reach further distances. However, most of them are, in some way, dependent on comparisons with Cepheid-determined distances.

Part II is devoted to providing brief accounts of several different methods:

Section 16, The Planetary Nebulae Luminosity Function (PNLF);

Section 17, The Tully–Fisher Relation (TFR);

Section 18, The Faber–Jackson Relation (F-J);

Section 19, Fundamental plane, the Dn-σ relation;

Section 20, Surface Brightness Fluctuation Method (SBF);

Section 21, Tip of the red giant branch (TRGB);

Section 22, Global Cluster Luminosity Function GCL.

Then, we take a short break (Section 23) to describe a discussion had by a group of experts at the end of the twentieth century on the merits of the distance measurement methods available at that time.

In Section 23, we describe the Cepheid calibration developments in the twentieth century, followed by the important method involving the use of Type Ia Supernovae (SN Ia) as standard candles in Section 24. This method led to the discovery that, contrary to the supposition at that time (end of twentieth century), the universe's expansion was not slowing down, but rather accelerating. There was a prevailing feeling that a cosmological constant was at work again in the field equations. Section 25 accounts for an alternative candle to measure the Hubble constant using Mira variable stars (Miras).

Section 26 briefly discusses the concept of a deceleration parameter ($q_0$) whose measurement provides values for the Hubble constant and the average density of matter in the universe. This method requires an accurate determination of the apparent luminosity of an object as a function of its redshift z. Then, we move on to the use of gravitational lensing to measure the Hubble constant.

To end with the description of the "late" route measurements for $H_0$, Sections 27–29 describe the lensing, masers, and the Sunyaev–Zeldovich Effect, respectively. Then, we move to an "early" method: The cosmic microwave background radiation (CMB) method, which is closely related to baryon acoustic oscillations (BAO) $H_0$ determination (Section 30). In Section 31, we briefly mention the standard siren method, which is a promising and different technique with which to determine the Hubble constant. In Section 32, we explain how shadows of supermassive black holes can be used to determine $H_0$, and in Section 33, we explain how fast radio bursts can also be used to constrain the Hubble constant.

In Section 34, we describe the current situation of the $H_0$ value. The results of "late" route measurements for $H_0$ all converge to a value within a very narrow range of uncertainty. The same can be said about the "early" route, but its values converge to a different $H_0$. Thus, the results of both ("early" and "late") sets of measurements do not coincide with each other within the error bars. We conclude this work by presenting some of the views and outlooks on this crisis, which have arisen because of the Hubble tension.

**Part I. Early historical roots.**

In the first section, we describe the historical events and personalities that played roles in the effort to determine astronomical distances in order to measure the Hubble constant.

## 2. The Great Debate of the 1920s

A little over a hundred years ago, on 26 April 1920, a debate was hosted by the US National Academy of Sciences. It took place in the Baird auditorium in the nation's capital. The place was an elegant, classically inspired room that featured a domed ceiling of Guastavino tiles. The audience was there to attend a day-long event that comprised an interesting debate. The debate was entitled "The distance scale of the Universe", with Harlow Shapley of Mt. Wilson Solar Observatory and Heber Doust Curtis of Lick Observatory as the debaters [1]. The first debater, Shapley, was a young astronomer who had previously practiced journalism, and the second, Curtis, a veteran astronomer better known for his comprehensive work on spiral nebulae for over a decade.

In addition to discussing the size and extent of the universe, which was the main reason for the debate, an attempt was also made to answer the query "Is the Milky Way an island Universe or is it just one of such galaxies"? Arguments were made in favor of the island universe by Shapley. For his part, Curtis championed the position that other galaxies exist alongside the Milky Way. The event began in the morning, with each of the astronomers' presentations addressing their own technical theses, and the actual debate took place later in the evening.

Shapley believed that the Milky Way contained all of the known and unknown celestial objects. In simple words, for him, the Milky Way was the whole universe. Shapley claimed that the evidence favored the island universe hypothesis, arguing that spiral nebulae (today identified as galaxies) were part of the Milky Way. He had estimated, using his own measurements, that the Milky Way was large (100 kpc). Shapley further argued that novae, which had been observed in spiral nebulae such as the Andromeda nebula, showed the same apparent brightness as those seen in the middle of the Milky Way [2]. Looking into former studies, he calculated that if Andromeda were not in the Milky Way, then it would be as large as 30,000 kpc., an implausibly-sized object for the audience present in the Baird auditorium. Consequently, according to Shapley, Andromeda was part of the island universe.

Perhaps the best argument presented by Shapley during the debate, and certainly one of the most forceful, was Adriaan van Maanen's measurements of the alleged rotation of the Pinwheel galaxy [3]. Van Maanen's measurements (now known to be incorrect) showed that the Pinwheel galaxy rotated on a time scale of years [4]. Shapley correctly argued that if Pinwheel were outside the confines of the Milky Way, then its spin rate would be greater than the speed of light, so it had to be concluded that such a "nebula" was also within the Milky Way.

In the course of the debate, an enigmatic issue concerning an astronomical observation made years earlier by Vesto Melvin Slipher received an easy answer from Shapley. We must recall that, some years before the debate, Vesto Melvin Slipher had measured the first wavelength shifts of spiral nebulae, meaning that they were recessing from us [5]. Regarding this, Shapley commented that Slipher's measurements meant that these objects were somehow repulsed away from the Milky Way's center by some unknown mechanism. For his part, Curtis did not comment on this matter.

In contrast, Heber Curtis was convinced that the Milky Way and Andromeda were independent galaxies, just two of many such bodies. On the subject of the size of the Milky Way, Curtis was of the opinion that, based on measurements of star counts in different regions of the sky, the Milky Way's diameter was only around 10,000 kpc. As for the Pinwheel galaxy argument, Curtis agreed that if the results of van Maanen were correct, Shapley was right. But Curtis rejected Adriaan van Maanen's results on the grounds that he considered van Maanen's results accurately unrealistic. Later astronomers have re-examined the measurements of Van Maanen, and have concluded that he made a serious error.

During the debate, Curtis noted that many novae on Andromeda were dim because their brightness was diminished by their distance from us. If one takes that distance into account, then the brightness of the observed novae approximately agrees with those that are closest to us in our Milky Way. Like Shapley, Curtis also had no explanation for the nebula recession observed by Slipher. Curtis pointed to evidence that the Milky Way had a spiral structure like any other nebula. Both Shapley's and Curtis' accounts were published a year later, in 1921 [6].

The debate ended a few years later when, as we will see below, Edwin Powell Hubble's work on Cepheid variables in several "nebulae" (today recognized as galaxies within the local group) settled the issue of the existence of external galaxies.

### 3. The First Candle

In 1908, Miss Henrietta Swan Leavitt was working at the Harvard College Observatory as a female "computer", as observatory assistants were familiarly called at that time. Formally, her position had the pompous name of "Curator of Astronomical Photographs". Leavitt had been hired some years prior by the observatory director, Edward Charles Pickering, to measure and register the brightness of stars. Her task was to catalog stars' luminosities by examining photographic plates from the observatory's collection. Her specific assignment was to identify variable stars. She utilized an instrument called a blink comparator that is no longer used today. By using that instrument, she would compare pairs of plates of the same star field taken a few days or weeks apart. This time-consuming operation involved manually flipping a screen back and forth quickly to suppress one image at a time, with the images in question on a pair of plates. In this way, a variable star would show up as a flashing spot. After analyzing plates of the Small Magellanic Cloud (SMC), taken in the lapse between the years 1893 and 1906, she compiled a catalog containing 1777 variable stars [7].

At some point, Leavitt conjectured that there was a relationship between the periodic luminosity changes observed in variable stars and their maximum brightnesses. She restricted her search to Cepheid variables that resided in the SMC. The reason for her opportune choice was her informed assumption that these stars must be at the same distance from Earth and, therefore, the comparison between their luminosities was well-founded. By 1912, Henrietta Swan Leavitt found that 25 Cepheid stars in the SMC would brighten and dim periodically [8]. Figure 1 shows the clear relationship that she found between the maximum (and minimum) brightness of each star and the length of its period, as she suspected.

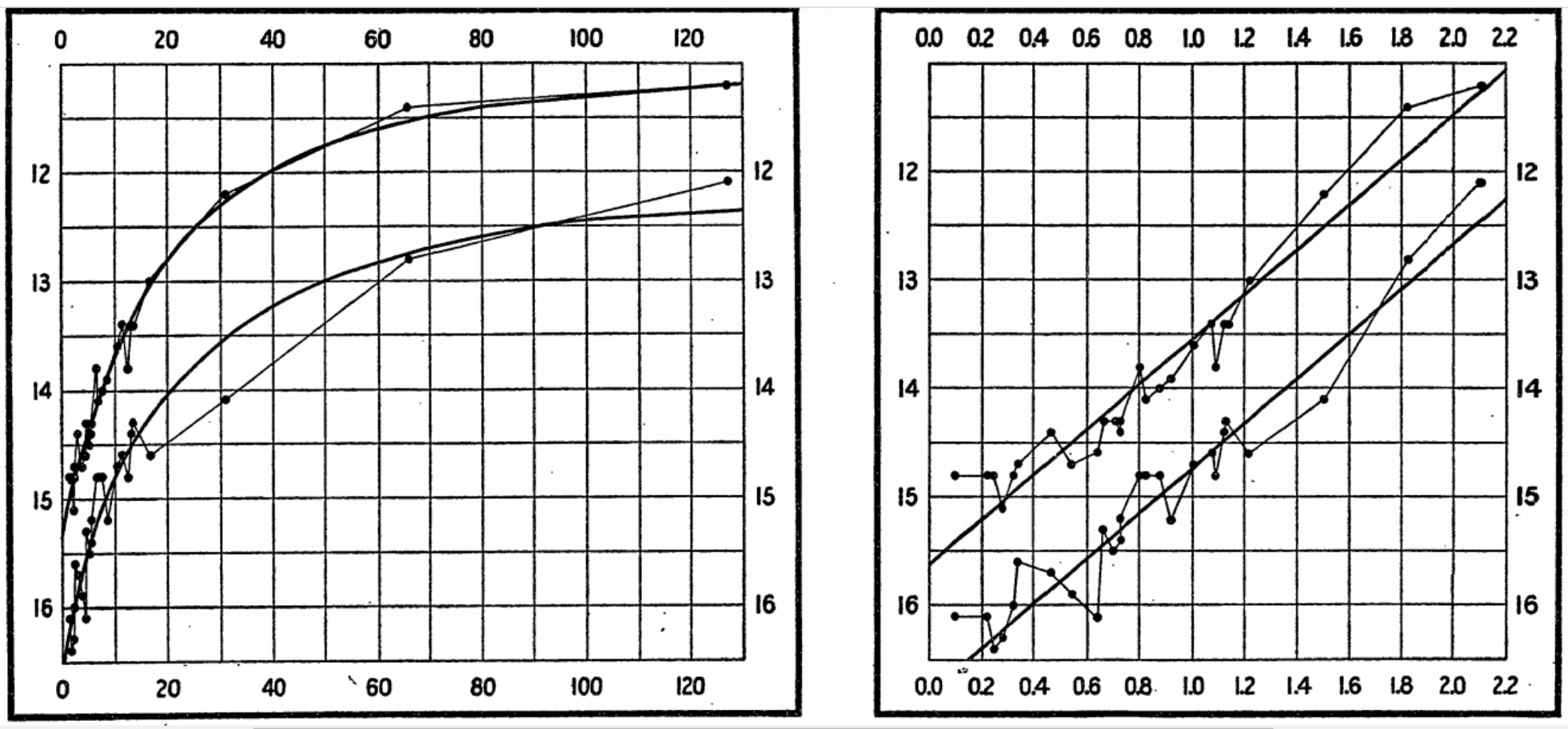

**Figure 1.** Period–luminosity curves and best fits. **Left**: Abscissas in days are equal to periods, ordinates corresponding to star magnitudes at their maxima and minima. **Right**: abscissas are equal to logarithms of the periods. Figure taken from ref. [8].

The discovery made by Miss Henrietta Leavitt led to the first standard-candle method for estimating distances to galaxies. In principle, the road to determining the actual distances to sky systems containing Cepheids was seeded. All that was needed was to find the distance to just one nearby Cepheid variable to calibrate Leavitt's period versus the luminosity relation (P-L relation).

Strangely, in her paper, Leavitt did not explicitly write out a mathematical relation. In modern notation, Leavitt Law is of the form $<M_v> = a + b \log_{10} P$, where P is the period of the luminous oscillation in days and *a* and *b* are constants to be determined. This pair of constants is univocally determined by a single point that astronomers call the zero point in their jargon. The zero point is defined as the absolute magnitude of a hypothetical Cepheid with P = 1 day.

Some people regret that Leavitt did not continue her research in determining stellar distances, but it must be remembered that her work at the observatory was limited to the study of stellar luminosities. In spite of this, at the end of her paper, Leavitt anticipated "It is to be hoped, also, that the parallaxes of some variables of this type may be measured".

The importance of performing these future measurements was not overlooked. A short time later, Ejnar Hertzsprung, by incorporating Leavitt's data, established the first period–luminosity calibration: $<Mv> = -0.6 - 2.1 \log_{10} P$. He also carried out a statistical parallax analysis on 13 Cepheids previously reported by Leavitt for which proper motions were available. Surprisingly, his paper had a misprint, as he reported a distance to SMC Cepheids of only 3000 light-years [9]. This result was well below the value of the actual distance, and perhaps this was the reason why his publication did not receive much attention. Some years later, Harold Shapley and Henry Norris Russell realized the error that had appeared in Hertzsprung's article and corrected the misprinted distance to 30,000 light years, see footnote 2 p. 434 in [10]. An interesting fact in Hertzsprung's work is that he ignored the interstellar absorption effects that cause dimming of light during its passage through the interstellar medium and alter the magnitudes of real stars.

Almost simultaneously to Hertzsprung's work, a very short note was published by Henry Norris Russell in 1913. In it, he estimates "the mean distance and real brightness… by the method of parallactic motion" of several variable stars, including maximum and minimum brightness estimates for the group of Cepheids previously reported by Leavitt (without citing Leavitt's publication) [11]. Some authors mistakenly indicate that Russell, like Hertzsprung, also presented a calibration of Leavitt's Law in this publication; however, this does not seem to be the case, as no such calibration appears in Russell's original paper. The point here is that Russell was among the first to perform an absolute magnitude determination of Cepheids. Yet, Russell did not, similarly to Hertzsprung, consider the effect of interstellar absorption.

However, it is interesting to note that in one of Russell's subsequent publications, this time in co-authorship with Harlow Shapley, absorption was considered [10]. The paper was concerned with the galactic distribution of eclipsing variables and Cepheids, and it came to the conclusion that there exists interstellar absorption of about two visual magnitudes per kpc, as was previously suggested by Edward S. King of the Harvard College observatory.

In effect, the need to consider interstellar absorption in luminosity measurements of stellar objects was implicitly suggested in 1912 by Edward S. King of the Harvard College observatory [12]. Since 1897, King had been involved in investigating how to obtain reliable luminosity measurements of stars in photographic plates [13]. For this purpose, he devised a photometric apparatus [14]. After applying his photometric methods to the moon and planets, he focused on determining the luminosity of stars following a protocol he carefully elaborated [15]. In connection with this task, King discussed the role of the absorbent medium in space and deduced evidence for its existence [16]. Then, he proposed an interstellar absorption of about 2 mag per Kpc [17]. A little later, in 1916, when he had already gathered more observations, he concluded "All indications point to the presence of an absorbing medium in space, or some factor which produces effects similar to absorption, by making the more distant stars redder" [18]. It is important to notice that this redness should not be misinterpreted as being produced by Doppler's effect. It is simply a light scattering effect caused by dust particles in interstellar space. A decade later, in 1927, King presented the hypothesis that a local cloud of absorbing matter, extending from the Sun to at least 100 light years, envelops our local star cluster [19].

Over the next two decades, there was a tendency for many astronomers to ignore the effects of interstellar dust absorption on stellar luminosities, which would ultimately influence the correct determination of Hubble's constant value using candle-based methods for several years. This may have been due to Shapley's change of mind from considering absorption to ignoring its effects. After the 1914 joint publication with Russel, Shapley was just beginning to determine distances using standard-candle methods. In 1915, Shapley published an article titled "Studies of Magnitudes in Clusters, 1. On the Absorption of Light in Space", where one of his conclusions read as follows: "It seems to be necessary to conclude that the selective extinction of light in space is entirely inappreciable, at least in the direction of the Hercules cluster" [20]. In this study, Shapley observed stars of all colors, so he correctly assumed that there was no absorption or redness of light on its path to Earth. In this case, there is indeed little absorption, since the studied clusters are far from the dust band that covers the plane of our galaxy. Therefore, the effects of absorption were imperceptible for Shapley. The problem came from extrapolating particular conclusions. It took him and many others several decades to abandon the trend of ignoring possible effects of interstellar absorption. This would have had a vast effect on the subsequent history of Hubble's constant.

## 4. The End of the Great Debate

In 1901, the wealthy heir Percival Lowell had spent part of his fortune establishing an astronomical observatory in Flagstaff, Arizona. Lowell was a highly controversial character, as he claimed to have observed a system of canals on Mars (the imaginary Schiaparelli's canals) that he believed criss-crossed the planet's surface, distributing water from the poles all over the red planet [21].

Around 1901, Lowell acquired an expensive state-of-the-art spectrograph for his observatory. In 1906, Lowell asked Vesto Melvin Slipher, one of the observatory's staff members, to survey spiral nebulae. The reason was that Lowell believed that the nebulae may have been solar systems in the process of formation. For this task, Slipher modified the spectrograph for nebular spectroscopy. After modifying the instrument, Slipher focused on the Andromeda Nebula. Not only did he record absorption lines, but he also saw that the lines were shifted toward the blue [22]. Interpreting these shifts as Doppler shifts, Slipher calculated that the Andromeda Nebula was hurtling at about $-300$ km s$^{-1}$ toward the Earth [23]. The fact that Slipher's discovery came from the premises of Lowell's observatory caused a stir of doubt regarding its soundness. Nevertheless, by 1914, Slipher had already collected data from 15 nebulae, of which 13 were receding and 2 were moving towards Earth [24].

In 1919, Edwin Powell Hubble joined the Mount Wilson observatory staff, where he had access to the 60-inch reflector as well as the just-completed 100-inch Hooker telescope. This instrument was by far the most powerful telescope in the world. In 1923, Hubble's research program became focused on locating novae in Andromeda using the Hooker telescope. By October 1923, Hubble had discovered what he took to be a nova in a nebula's outer edges. But as Hubble examined previously obtained plates of the same region, he noticed that his newly discovered "nova" regularly exhibited brightness changes, so upon constructing a light curve for its varying brightness, he realized that what he had found was a Cepheid variable, not a nova.

Hubble measured the period and apparent brightness of the Cepheid. Then, by employing Leavitt's Law, he arrived at a distance of around 900,000 light years (estimated to be approximately 2.5 million light years presently), placing Andromeda Nebula well outside the limits of Shapley's estimate of the Milky Way's size. In Hubble's best self-promoting style, his finding was first released by the press [25]. Hubble's formal announcement, entitled "Extragalactic Nature of Spiral Nebulae," was delivered in absentia by Henry Norris Russell to a joint meeting of the American Astronomical Society and the American Association for the Advancement of Science which was held at the end of December 1924 [26]. Thus, in the words of Shapley's opponent in the Great Debate, Heber D. Curtis, it was not until 1924 that "… all doubts as to the island Universe character of the spirals were finally swept away by Hubble's discovery of a Cepheid" [27].

Hubble's results for Andromeda were not formally published in a peer-reviewed scientific journal until 1929 [28], but "The Great Debate" was over in 1924.

However, a delicate conflict had endured at Mount Wilson between Hubble and his observatory colleague Adriaan Van Maanen. Recall that it was Van Maanen's measurements that Shapley used to argue that the spiral nebulae were located within the limits of the Milky Way. This conflict between these two characters (this time, Hubble and Van Maanen), turned into a public dispute, which ended in 1935 with the publication of a brief note by each of them [29,30]. In Van Maanen's note, he conceded that his measurements should be taken with caution.

## 5. Velocity–Distance Early Searches

After Hubble found that the spiral nebulae were actually galaxies like ours, Slipher's findings attracted the attention of some astronomers. On the one hand, most nebulae (now known to be galaxies) were rapidly moving away from Earth; on the other, a few were approaching it. Why did this imbalance exist? What was causing it?

In 1917, Willem de Sitter, a Dutch mathematician from Leiden University, had discovered a solution to Einstein's field equations of general relativity [31]. De Sitter's solution for an empty universe had very important cosmological consequences: specifically, an exponentially expanding universe. This signified that, if the implications of De Sitter's model were true, the redshifts observed by Slipher were indeed reflecting the predicted expansion of the universe. Whether it was the result of De Sitter or Slipher's findings, the fact is that some astronomers undertook the challenge to search for some relationship between redshifts and the distances of those light-emitting objects. General relativity theory, as interpreted by De Sitter, suggested that this relation should exist.

One of the first astronomers to accept the challenge was Carl Wilhelm Wirtz. His observations produced correlations between radial velocities and nebular distance indicators [32]. As distance indicators, he used comparisons between the diameters of galaxies. He was the first to attempt the search for a relation between radial velocity and distance. As a result, he found $V$ (km/sec) = $2200 - 1200 \log (D_m)$, where $D_m$ represents the angular diameter (scales like $1/r$) in arc minutes of the observed nebula. Clearly, this was incorrect, but his relation followed the right tendency for V as it increases with distance r.

Another scientist undertaking the challenge of finding a velocity–distance relation, albeit for different reasons, was Ludwik Silberstein (Einstein's antagonist [33]). Silberstein attempt to establish a velocity–distance relation was intended to determine the curvature of the universe, which he considered to be fixed. He used De Sitter's results to derive a formula for the shift of the spectral lines emitted by stars [34]. For distant stars, Silberstein found that, with the limit of small velocities, the Doppler shift is $\Delta\lambda/\lambda = \pm r/R$, where R is the radius of curvature of the universe and $\Delta\lambda$ is the shift in the wavelength $\lambda$ of the line. At this point, it is convenient to remember that today, a redshift value "z" is reported ($z = (\lambda_{obs}/\lambda_{em}) - 1$, where $\lambda_{obs}$ is the observed and $\lambda_{em}$ is the emitted wavelength). Silberstein applied his formula to a list of stellar clusters as well as to the small and large Magellan clouds. The application of this law to this mix of objects (some approaching and others receding) gave him a value for the radius of curvature of the universe of R of about $10^8$ lyr. As expected, it did not take long for Silberstein's work to be criticized by various astronomers, including Arthur Eddington [35].

One earlier attempt to obtain a velocity–distance relationship was that of the Swedish Knut Emil Lundmark. In 1924, he plotted the radial velocity of 44 galaxies against their estimated distances [36]. He assumed that Andromeda was 200,000 pc away. Then, he made rough determinations of the distances to other galaxies by comparing their angular sizes and brightnesses to that of Andromeda. Figure 2 shows Lundmark's plot. He concluded that there may be a relationship between galactic redshifts and distances, but "not a very definite one".

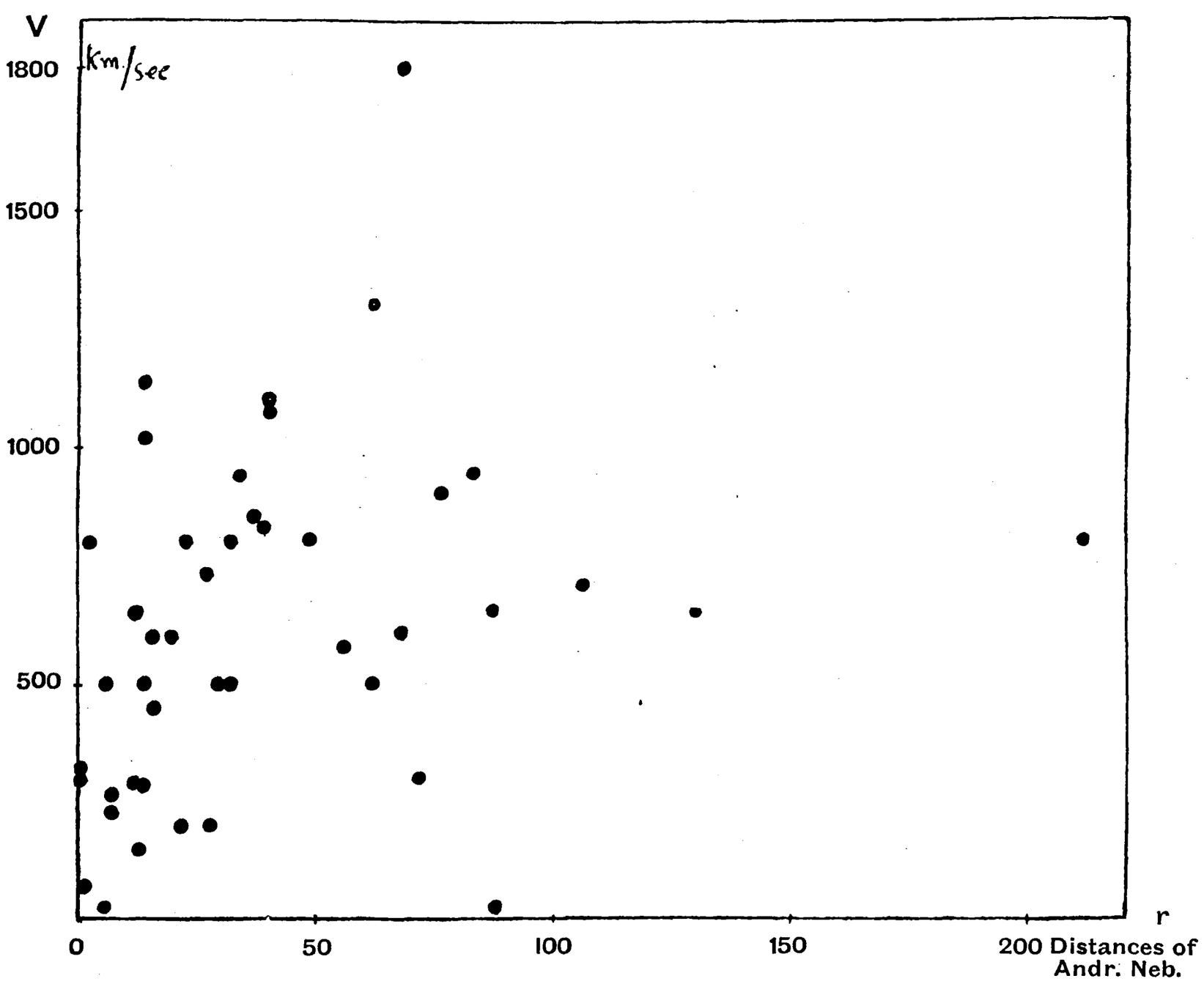

**Figure 2.** Knut Lundmark's 1924 [36] relation between relative distances and observed radial velocities of spiral nebulae (The scale unit is the distance to the Andromeda "nebula").

By the mid-1920s, there were no convincing studies on a possible relationship between recession velocity and distance. The collected evidence was far from being able to test De Sitter's model. The chief issue involved the imprecise estimates of distances to galaxies. It was then that Hubble decided to involve himself in the problem.

## 6. Hubble's Entrance

Since entering Mount Wilson, Hubble had made good use of the 100-inch Hooker telescope. By the early 1920s, he had already detected 11 Cepheid variables in Barnard's Galaxy (NGC 6822) and derived an estimate for its distance [37]. Hubble's detection was a milestone in astronomy, as it was the first system beyond the Magellanic clouds to have its distance determined.

However, the obtained value of the distance was well below the currently accepted value, because Hubble used what was then the most recent calibration of Leavitt's Law, made by Shapley in 1925. Today it is known that the usage of Shapley's calibration underestimated distances. Subsequently, by using the same Cepheid method and comparing the mean luminosity between galaxies (a kind of "embryonic ladder"), he continued with the determination of distances to other galaxies, such as Andromeda and the Triangulum Galaxies. Step by step, Hubble and colleagues piled up estimates on galactic distances.

By 1929, the spectral shifts for 46 galaxies had been already measured, the great majority by Vesto Slipher. In that same year, Hubble plotted the recession velocity (which, for nearby objects, is given by $v = c \cdot z$) versus nebular distances that he considered to be fairly reliable, i.e., most his own estimates plus two by Shapley and four by Humason. Of the 22 redshift values plotted by Hubble, 18 had previously been measured by Slipher. Hubble did not reference Slipher in his publication [38].

Hubble asserted that his plot of redshift versus distance was, at least to the first approximation, best represented by a linear relation. Curiously, instead of expressing the relation into the form that would soon become typical ($v = c \cdot z = H \cdot r$, with H as a constant and following what would later be termed the Hubble–Lemaitre law), Hubble wrote it in terms of the solar motion equations. Nevertheless, by interpreting redshifts as Doppler shifts, what Hubble found was a simple linear relation between distances to galaxies and their recession speeds (as determined by their redshifted spectral lines). Final-ly, in his paper, Hubble announced his intention to expand his research and made a concluding remark:

"In order to investigate the matter on a much larger scale, Mr. Humason at Mount Wilson has initiated a program of determining velocities of the most distant nebulae that can be observed with confidence. The outstanding feature, however, is the possibility that the velocity distance relation may represent the De Sitter effect and hence that numerical data may be introduced into discussions of the general curvature of space".

Indeed, as he stated in his 1929 article, in order to extend his research, Hubble enlisted the help of Milton La Salle Humason, a colleague from Mount Wilson who had toured the entire hierarchy of the observatory, from mule skinner in the early days of the establishment through janitor until his curiosity led him to learn from other astronomers and, eventually, become one of them.

Humason focused his instruments on the very faint and presumably farthest galaxies. Thanks to his ability, exposure times of plates taken by him were adequate to allow him to identify ionized Calcium H and K spectral lines, and short enough to prevent entire spectra from becoming continuous. For each of the observed galaxies, Humason calculated the z values of their shift and distances. In March 1931, they presented their extended results [39] (see Figure 3). The value for the H constant which they found for Hubble's empirical relation was about 500 km/s/Mpc.

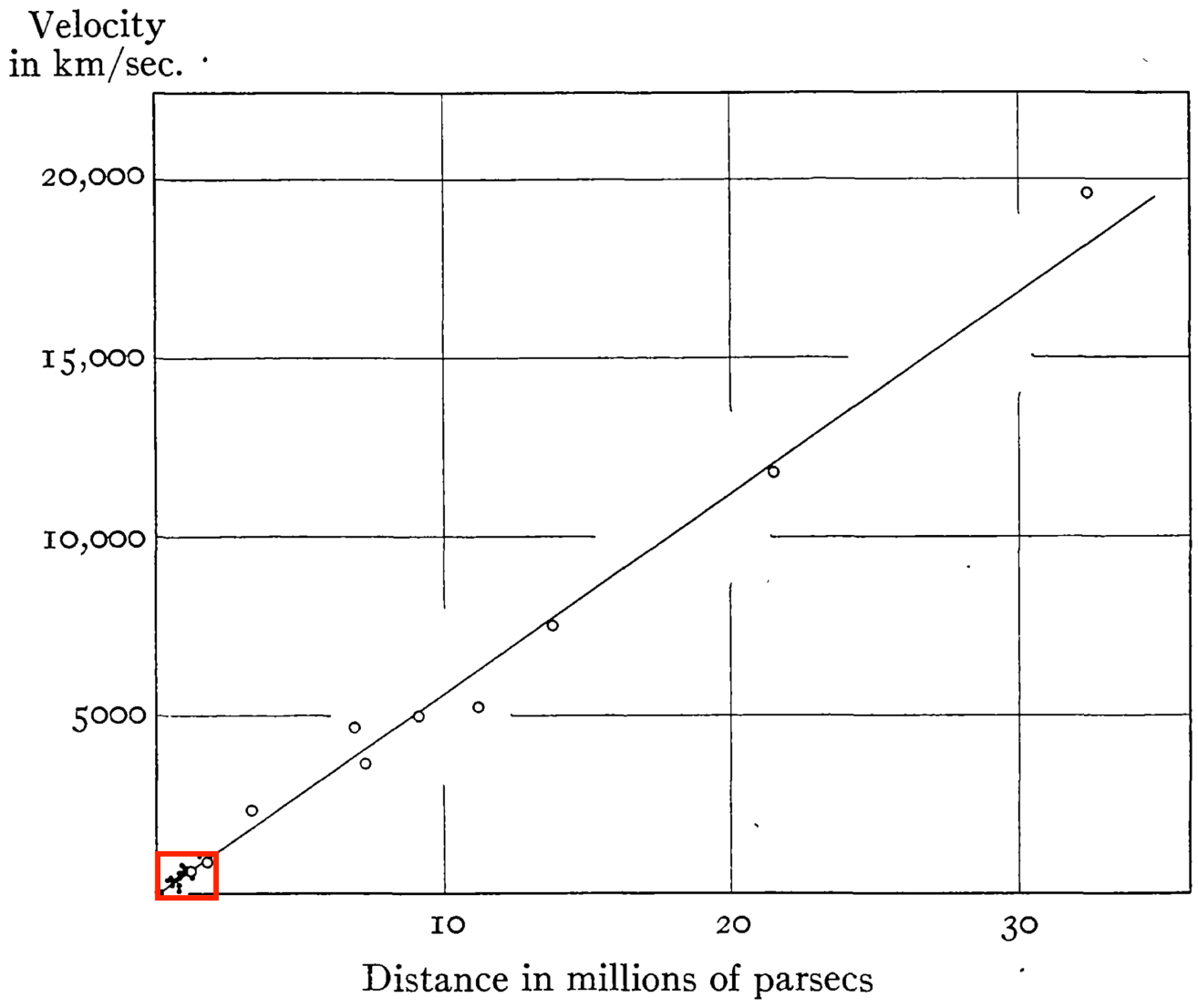

**Figure 3.** Galactic redshift vs. distance, plotted by Hubble and Humason (1931) [39]; the rectangle in the lower left corner encloses data points plotted in 1929.

## 7. The Emergence of Cosmological Relativity

Reactions to the appearance of general relativity by the broad spectrum of scientists showed widespread indifference among the less informed. Take, for example, a fragment of the correspondence (1919) between the distinguished astronomer George Ellery Hale and the then-Assistant Secretary at the Smithsonian Institution, Charles Greely Abbot. This epistolary exchange took place a year before the so-called "great debate", about which we have already written in a previous section. In a first letter, Hale had suggested two topics for Abbot to choose from for the keynote debate at the National Academy of Science (NAS) meeting the next following year (April 1920). The suggested topics were general relativity and the size of the Milky Way. Hale, in fact, favored relativity, but Abbot had the last word on this choice. Abbot's response illustrates our point:

> *"As to relativity, I must confess that I would rather have a subject in which there would be a half dozen members of the Academy competent enough to understand at least a few words of what the speakers were saying if we had a symposium upon it. I pray to God that the progress of science will send relativity to some region of space beyond the fourth dimension, from whence it may never return to plague us"* [40].

In fact, the plague had started earlier and was there to stay.

After completing his general theory of relativity in 1915, one of the first things Einstein attempted was to apply it to model the universe at large scale. Einstein's idea of the universe was that of a static entity, an idea shared by the majority of scientists of that time, and there was no known reason at that time for Einstein to doubt it [1].

To ensure the stability of his static universe, Einstein introduced a change in his field equations by adding what he called a "cosmical" term proportional to the metric tensor that he thought would guarantee the large-scale immobility of his model universe. This additional term gave rise, even in empty space, to a repulsive force which would allow his model universe to remain static, counterbalancing the gravitational attraction of the matter within it.

In 1917, Einstein published his renewed version of the field equations, adding the mentioned "cosmical" term to his original field equations. He solved his field equations considering an isotropic and homogeneous cylindrical universe, that is: space dimensions corresponded to a sphere, but the time dimension was uncurved. His paper was titled "Cosmological considerations in the General Theory of Relativity". Thus, his renewed field equations written in standard form were as follows ("λ" is today; "Λ" is the cosmological constant [2]),

$$R_{\mu\nu} - \frac{1}{2} g_{\mu\nu} R - \lambda g_{\mu\nu} = -\kappa T_{\mu\nu},$$

This calculation shows that $\lambda$ is proportional to $\rho$ , the mean mass density of the universe, and inversely proportional to $R$, its radius of curvature, i.e., $\lambda = 1/R^2 = \frac{1}{2}\kappa\rho c^2$. Regarding this result, Eddington readily pointed out an issue with the marginal stability of Einstein's universe model when he wrote "the question at once arises, by what mechanism can the value of $\lambda$, be adjusted to correspond with M"? In Eddington's book, M is the total mass of the universe [41]. With this comment, Eddington hinted that Einstein's model universe was static, but unstable. In addition, Einstein's world model offered no explanation of the observed redshifts. Nonetheless, it was the first step that marked the beginning of modern cosmology as the scientific study of the origin and structure of the universe. The second step was made by a Dutch mathematician and astronomer named Willem De Sitter.

Beginning in 1911, De Sitter maintained regular correspondence with Arthur Eddington on matters of astronomical interest. De Sitter held the chair of astronomy at Leiden University in the Netherlands, where he also discussed issues related to the theory of relativity and its consequences with his university colleagues, Hendrik A. Lorentz and Paul Ehrenfest. He had also met sporadically with Einstein for the same reasons. De Sitter had contributed to spreading relativity in the English-speaking world, since the Great War that prevailed in those days prevented the free diffusion of Einstein's ideas from Germany. In this regard, in 1916, he published two articles entitled "On Einstein's theory of gravitation and its astronomical consequences, first part and second part", which contributed to the diffusion of relativity [42,43].

In 1916, Paul Erhenfest suggested to De Sitter that some of the difficult problems associated with the idea of an infinite universe could be avoided by assuming a closed universe. It was in 1917, upon learning of Einstein's publication, that De sitter produced three solutions for the field equations, retaining the "cosmical" term. He stipulated that his model should be isotropic and static. He constrained the spatial part of space–time to constant curvature. One of the models was Einstein's own. It contained matter related to the "cosmical" term, but zero pressure. Another contained zero pressure, zero matter, and no "cosmical" term. The last is what we know now as De Sitter's model [44]. Density and pressure were both zero, but he retained the "cosmical" term. As this term was introduced by Einstein to counterbalance the gravitational contraction, and since De Sitter did not include mass, his universe expanded. If some light emitter was found within this universe, what would happen is that if the universe were to expand, a redshift of its light would be observed. This phenomenon was labeled the De Sitter effect in its time.

Despite the fact that the model of the universe proposed by De Sitter was quite far from what would be considered a real universe, the model was very relevant because it, at least, predicted the presence of redshifts.

De Sitter's solutions to Einstein's field equations caught the attention of several theorists among them—Ludwick Silberstein, Hermann Weyl, and Richard Tolman—who began to explore the physical aspects of the given solutions. It was around this time that two individuals unknown to experts in the field independently made great strides in solving the field equations. The first was a young Russian man named Aleksandr Aleksandrovich Friedmann, and the second was a Belgian Catholic priest, Georges Henry Joseph Édouard Lemaître.

**8. Friedmann**

In 1922, a short note by Einstein of only eleven lines appeared in the Zeitschrift für Physik journal [45]. It reads, "*The results contained in the cited work regarding a non-stationary world seemed suspicious to me. In fact, it turns out that this given solution is not compatible with the field equations*". In this belittling note, Einstein was referring to a paper by A. Friedmann that had appeared that same year [46].

At that time, Alexander Friedmann held an academic position at the University of Saint Petersburg (formerly Petrograd and renamed Leningrad). Friedman's situation at the end of the Russian civil war must have been difficult, as he was academically isolated from the rest of the world. News about astronomical discoveries like those of the redshift should not have flowed easily at this time, when there were still skirmishes between the red and the overpowered white armies.

In the paper in question, which was belittled by Einstein, Friedmann gave several valid solutions to Einstein's equations with the "cosmic" $\lambda$-term, assuming, as Einstein did, a homogeneous, isotropic, and positively curved universe. The main difference that Friedmann introduced, in contrast to Einstein's and De Sitter's models, was allowing the radius of curvature $R$ to vary with time. Friedmann wrote the line element $ds^2 = g_{ik}dx^i dx^k$ as:

$$ds^2 = \mathrm{R}^2\left(dx_1^2 + sin^2 x_1 dx_2^2 + sin^2 x_1 sin^2 x_2 dx_3^2\right) + M^2 dx_4^2$$

where $M = 2\,\pi^2 \rho R^3$ stands for the mass of the universe, $\rho$ its density, and $R$ a distance scale proportional to the curvature of space.

In this 1922 paper, Friedmann begins by dealing with static cases where R is not a function of time $(x_4)$. Then, he reduces his model to the static Einstein and Sitter universes by setting the function $R$ as equal to $R^2/c$, where the value $R$, i.e., the radius of curvature, is constant. To recover Einstein's and De Sitter's models, he set $M = 1$ for Einstein's and $M = \cos(x_1)$ for De Sitter's, respectively. He demonstrated that these are the only two solutions for static universes. Then, he moved on to the non-static cases.

Using Einstein's fundamental equations and retaining $\lambda$, Friedmann then derived equations for the determination of $R = R(x_4)$ and the matter density $\rho = \rho\ (x_4)$, both of which were allowed to vary with time according to Einstein's equations:

$$R_{ik} - \frac{1}{2}\, g_{ik} R + \lambda\, g_{ik} = \kappa\, T_{ik}$$

Neglecting pressure terms, the time-dependent $R(x_4)$ is determined by $\lambda$ and $\rho$. The energy–momentum tensor reduces to

$$T_{44} = c^2 \rho g_{44} \quad T_{4i} = 0 \quad T_{ik} = 0 \quad for\ i,k = 1,2,3$$

Setting $i = k = 1,2,3$, Friedmann obtained his equation for R:

$$\frac{\dot{\mathrm{R}}^2}{\mathrm{R}^2} + \frac{2\mathrm{R}\ddot{\mathrm{R}}}{\mathrm{R}^2} + \frac{c^2}{\mathrm{R}^2} - \lambda = 0,$$

where a dot stands for the time derivative. Setting $i = k = 4$, he obtained the following relationship:

$$\frac{3\dot{R}^2}{R^2} + \frac{3c^2}{R^2} - \lambda = \kappa c^2 \rho,$$

which is known as Friedmann equation today. Since the value of $\lambda$ was unknown, he considered different possibilities and found that, depending on the ratio between the attraction of the gravitational force due to the total mass of the universe and the "cosmic" term, one can obtain universes that expand or contract or vary periodically as a function of time. In a second paper in Zeitschrift fuer Physik, and in a book published in 1923, he explored the open-geometry structure of the universe [47,48].

Friedmann speculated on the age, mass, and size of some of his universes, but did not make comparisons with astronomical data known at the time because he thought that the uncertainty of these values did not warrant it.

He also did not incorporate the redshifts measured by Slipher, perhaps because he ignored their existence. Friedmann died of typhoid fever in 1925 at age 37. His work did not attract attention and was erased from memory until it was revived years later by Eddington.

As we have already mentioned, at the time Friedmann sent his paper for publication, he held an academic position at the University of Saint Petersburg (Petrograd). One of his colleagues at the university, Ivan Kruthoff, had attended a meeting at Leiden in May 1923 where Einstein was also staying (at Paul Ehrenfest's home). Einstein frequently visited Leiden, as he had the appointment of special professor ("Bijzonder Hooglerаar"), a position promoted by Ehrenfest [49]. Friedmann saw Kruthoff's visit to Leiden as an opportunity to discuss his previously delivered letter to Einstein, in which he presumably provided more details about his solutions. In this way, he ensured that Einstein would read his letter. We must mention that Friedmann and Ehrenfest maintained correspondence, since the former had been an outstanding student of Ehrenfest when he was a lecturer at Saint Petersburg in 1912.

When Einstein realized that the solutions found by Friedmann were mathematically "correct and clarifying", he wrote a short note rectifying the error of appreciation he had made in his previous note, but warning that they were non-static solutions. "They show that in addition to the static solutions to the field equations there are time varying solutions" [50].

## 9. Lemaître

On Friday, 10 January 1930, at Burlington House, Piccadilly, an important meeting of the Royal Astronomical Society was held. Willem De Sitter gave a lecture on the discovery Hubble had just recently announced: the probable existence of a linear relationship between the recession velocity of spiral nebulae and their distances [51]. De Sitter explained in detail to the audience how the distances to the spirals were determined. He also admitted that it was difficult to reconcile this result with one of the two universe theories, the one proposed by him and the other by Einstein.

During the discussion that followed the talk, Eddington wondered why there were only two static solutions. One of which was Einstein's, and in the other (De Sitter's), the static universe expanded as soon as matter was introduced into it. To Eddington's concern, De Sitter commented that it would be desirable to investigate how to include matter into his model, but he admitted that the difficulty that would arise would be keeping the universe static. However, Eddington conceded that, perhaps, it would not hurt to investigate non-static intermediary solutions. From the discussion that followed the talk, the idea of a static universe prevailed, as it was ingrained in the minds of academics despite Hubble's results. The résumé of the discussion that took place at the meeting was published in the February issue of *The Observatory* [52].

A few weeks after the meeting at Burlington House, Eddington received a letter from the Belgian Jesuit Georges Lemaître which stunned him. But before reproducing part of the letter, it is timely to recall some earlier undertakings of Lemaître.

In the past, from 1923 to 1924, the Belgian Jesuit had spent a season of research on general relativity in the U.K., at Cambridge. There, he took lectures by Eddington on "Relativity Theory of Electrons and Protons" and "Fundamental Theory", giving him the opportunity to advance his knowledge on the subject [53]. As a visiting student, Lemaître produced a paper in which he generalized the definition of simultaneity [54].

After his stay at the English university, Lemaître and Eddington crossed the Atlantic to participate in the 94th annual meeting of the British Association for the Advancement of Science in Toronto (6–13 August 1924), where Eddington spoke on relativity and the bending of starlight. Then, in September 1924, Lemaître arrived at Harvard College Observatory to work on Cepheids under Harlow Shapley's supervision. Lemaître spent 9 months there, where he wrote a paper criticizing De Sitter's universe. The Belgian priest spotted that the geometrical coordinates chosen by De Sitter in his model introduced a spurious inhomogeneity. To be precise, what Lemaître found was that De Sitter's model produced a privileged point with properties different from the others, a fact which violated the principle of spatial homogeneity. De Sitter chose his line element as:

$$ds^2 = \mathrm{R}^2\left[-d\chi^2 - sin^2\chi\left(d\theta^2 + sin^2\theta d\phi^2\right) + cos^2\chi d\tau^2\right],$$

where $R$ is the constant radius of the universe with coordinates: $\chi, \theta, \phi, \tau$. The spatial part of the line element is constant and time-independent, so it produces a static universe whereas the temporal part, $\mathrm{R}^2 cos^2\chi d\tau^2$, depends on the spatial coordinate $\chi$, except when $\chi = 0$, thus distinguishing this point from others. Lemaître noted that "*It is clear that such an introduction of an apparent center in a Universe which, by definition, has none is objectionable for a study of the properties of this Universe*". That is, De Sitter's model contravened his initial premise of spatial homogeneity. Lemaître published a paper on the subject, proposing a new choice coordinates that avoided the aforementioned problem [55,56].

In late 1924, the Belgian priest had the opportunity to attend the 33rd meeting of the American Astronomical Society in Washington, D.C., where he heard Henry N. Russell announcing Hubble's theory on the distance to Andromeda using the Cepheid observations, a presentation that, as we have already pointed out, was of celebrated consequence as it ended the "Great Debate". Russell also read a paper on "Stellar Evolution" by Eddington, who had already returned to England.

Afterwards, on 18 June 1925, for the purpose of learning more about extragalactic distances, Lemaître traveled to California to meet Hubble in person at Caltech. On his way back to the East Coast, Lemaître stopped at the Lowell Observatory in Arizona to visit Vesto Slipher in order find out more about the method of measuring redshifts. All of these visits allowed him, a theoretician, to become familiar with astronomical observations.

On June 1925, Father Lemaître paid a visit to the Old Continent to attend the second triennial meeting of the International Astronomical Union (14 June to 22 July) at Cambridge, England. During the meeting, Slipher showed his latest redshift measurements, Hubble discussed his proposal on galaxy classification based on morphology, and De Sitter received an honorary degree of Sc. D. from Cambridge [57].

During the meeting, a delicate question was scheduled for discussion. This was the admission of Germany and the Central powers to the International Astronomical Union. At that time, the wounds caused by the First World War were just healing. However, the decision to admit Germany to the Club was deferred until after participants had had the opportunity to discuss the matter informally. Having previously foreseen this situation, the organizers of the event had planned a large series of social activities (visits, banquets, and garden parties) that would allow for the exchange of opinions among the attendees on the aforementioned matter and, of course, on scientific matters [58]. At those events, Lemaître had had the opportunity to renew acquaintances and perhaps to assess the "*air-du-temps*" on a non-static model of the universe.

Back in Belgium in 1927, he learned by mail that MIT had conferred him a doctoral degree (Ph.D.) for a thesis on general relativity [59], and he was exempt from "*viva voce*" (oral defense). Upon returning to Louvain, he became part of the academic staff at the Catholic University. There, in 1927, he finished writing his paper on a non-static model of the universe, which he published in the Annals of the Scientific Society of Brussels in French.

In autumn 1927, Einstein was in Brussels on the occasion of the Fifth Solvay International Conference. The congress met then to discuss the newly formulated quantum theory. Obviously, in those moments, Einstein's thoughts were not centered on cosmology. The new formulation of quantum mechanics made him feel uneasy.

For Father Lemaître, this was an opportunity to exchange ideas directly with the "Pope of Relativity". He took a train from Louvain to Brussels. He himself later described this meeting [60]: "While walking in the alleys of Leopold Park, [Einstein] told me about an article, little noticed, that I had written the previous year on the expansion of the Universe and that a friend [Auguste Piccard who was also present during the stroll] had made him read. After some favorable technical remarks, he concludes by saying that from the physical point of view it seemed to him completely abominable (*tout à fait abominable*)". During that encounter, Lemaître also learned from Einstein of Friedmann's previous work published in 1922. Lemaître did not know German, which may explain why he had ignored the existence of Friedmann's paper. Later that day, during a ride in a taxi with Einstein and Picard, Lemaître developed the impression that "Einstein was hardly aware of the astronomical facts" [61]. Einstein's comment ("*magister dixit*") must have surprised Lemaître, and in spite of his Einstein's "*anathema*", Lemaître was tenacious with his ideas.

Before meeting Einstein, Father Lemaître had already sent his manuscript to Eddington soon after receiving reprints of his paper, but received no response; his former mentor classified the manuscript without really reading it. In July 1928, Lemaître traveled to Leiden, where De Sitter chaired the third assembly of the International Astronomical Union. Unfortunately, Lemaître did not have a chance to discuss this matter with him.

As mentioned above, the report from the January 10th meeting at Burlington House reproducing the discussion between Eddington and De Sitter regarding the possibility of finding non-static intermediate solutions to Einstein's field equations was published in 1930 in *The Observatory*. As soon as an issue reached Lemaître, he immediately wrote a letter to Eddington attaching copies of his 1927 paper and asking Eddington to send a copy to De Sitter [62] (A similar solution had previously been given by A. Friedman [63]):

> "Dear Professor Eddington, I just read the February No., of the Observatory and your suggestion of investigating of non-statical intermediary solutions between those of Einstein and de Sitter. I made these investigations two years ago. I consider a Universe of curvature constant in space but increasing in time. And I emphasize the existence of a solution in which the motion of the nebulae is always a receding one from time minus infinity to plus infinity".

From reading the full content of Lemaître's missive, Eddington also learned that Alexander Friedmann had produced (years before, in 1922) a non-stationary solution to Einstein's field equations. Friedmann's solution was like the independently rediscovered one, which was attached to the letter sent by Lemaître in 1927. The letter also indicated that, two years prior, the Jesuit had sent De Sitter a copy and lamented that he likely did not read it [64].

Upon learning of this, Eddington warned De Sitter. The latter was, at that moment, writing a manuscript on the same topic, focusing on the astronomical consequences of relativity. De Sitter, realizing the great importance of the solution, modified his article to include and comment on Friedmann's and Lemaître's accomplishments [65].

In the first part of his paper, De Sitter argues why "static solutions to field equations must be rejected and that the true solution represented in nature must be a dynamical solution". Then, he continues: "A dynamical solution [to the field equations] … is given by Dr. G. Lemaitre which had unfortunately escaped my notice until my attention was called to it by Professor Eddington a few weeks ago in a paper published in 1927". Next, he goes over the solution for English-speaking readers. At the end of his publication, De Sitter praises Lemaître's achievement.

For his part, Eddington published an article in which he discussed the instability of Einstein's model and also applauded the Belgian priest: "…we learnt of a paper by Abbé G. Lemaître which gives a remarkably complete solution of the various questions con-nected with the Einstein and De Sitter cosmogonies" "…my original hope of contributing some definitely new result has been forestalled by Lemaître's brilliant solution" [66].

Eddington embraced Lemaître's model with great enthusiasm, distributing it to colleagues and arranging for it to be translated into English and republished in the Monthly Notices of the Royal Astronomical Society. We shall comment on a mathematical formula that disappeared in the translation, but first, we will describe the contents of Lemaître's celebrated paper.

## 10. Lemaître's Expanding Universe

Lemaître's paper begins by considering an Einstein universe, where the radius is allowed to vary in an arbitrary way. Then, he restricts himself to those solutions in which the three-dimensional space has complete spherical symmetry [3] and is filled with matter comparable to a rarified gas, that is, uniformly and homogeneously distributed through space where its molecules are the extragalactic nebulae. Hence, the corresponding total density of energy in matter $\delta(t)$ depends on time ($\delta(t)$ is the trace of the energy momentum). For his line element, he considers:

$$ds^2 = -\,\mathrm{R}(t)^2\, d\sigma^2 + dt^2,$$

where $\mathrm{R}(t)$ is the radio of curvature of the 3D space and $\sigma$ is the spatial volume element.

Then, he uses some lines to explain why he considers the matter's contribution to pressure to be negligible, but not $p$, the radiation pressure. Thus, he denotes the total energy density $\rho$ as $\rho = \delta + 3p$ . Also, he assumes a closed universe, so energy is conserved. Under these assumptions, the expression for the conservation of energy turns out to be:

$$\frac{d\rho}{dt} + \frac{3\dot{\mathrm{R}}}{\mathrm{R}}(\rho + p) = 0$$

The volume of space is $V = (4\pi/3)\, R^3$, so energy conservation becomes $d(V\rho) + p\,dV = 0$. This means that the work done by radiation pressure plus the variation in total energy add up to zero. Then, he explains why pressure coming from the mass loss of stars that is converted or transformed into energy does not contribute to the pressure. Also, he recovers Einstein's model by setting the condition of the constancy of the universe's radius while setting $\rho = 0$, as well as retrieving De Sitter's solution.

$$\frac{3\dot{R}^2}{R^2} + \frac{3c^2}{R^2} - \lambda = \kappa c^2 \rho$$

Then, Lemaître turns his attention to Doppler's effect. Here, Lemaître explains that the cosmological nature of spectral shifts is not due to relative motion between the observed object and the observer, but to the variation of the radius of the universe. He notes that the large recession velocities of extragalactic nebulae are a cosmological effect due to the fast expansion of the universe, perhaps due to the pressure of radiation. Next, he directs his interest to the lightshift.

For a light ray emitted at space coordinates $\sigma_1$ and observed at $\sigma_2$, and since it follows a geodesic ($ds = 0$) from Lemaître's line element, the equation for a light ray is:

$$\sigma_2 - \sigma_1 = \int_{t_1}^{t_2} \frac{dt}{\mathrm{R}}$$

If the light ray emission in $\sigma_1$ occurs at time $t_1 + dt_1$, where $dt_1$ is the period of the emitted light and is observed in $\sigma_2$ at $t_2 + dt_2$, where $dt_2$ is the period of the observed light, then the above integral implies:

$$\frac{dt_2}{\mathrm{R}_2} - \frac{dt_1}{\mathrm{R}_1} = 0, \quad \frac{dt_2}{dt_1} - 1 = \frac{\mathrm{R}_2}{\mathrm{R}_1} - 1.$$

where $\mathrm{R}_1$ and $\mathrm{R}_2$ the values of R at $t_1$ and $t_2$. Since wave length $\lambda$ is related to wave period $dt$ by $\lambda = c\,dt$, then light emitted with wavelength $\lambda_1$ at $\sigma_1$ will arrive at $\sigma_2$ with a wavelength

$$\lambda_2 = \lambda_1 \frac{R_2}{R_1}$$

$\lambda_1$ is subtracted from both sides, and the equation is rearranged:

$$\frac{\lambda_2 - \lambda_1}{\lambda_1} = \left(\frac{R_2}{R_1} - 1\right) = \frac{\Delta\lambda}{\lambda_1}$$

If the redshift $\Delta\lambda/\lambda$ is interpreted as being due to Doppler's effect due to velocity $v$, for $v << c$, $\Delta\lambda/\lambda = v/c$. Thus, we can assume that $\lambda_2 - \lambda_1 = \Delta\lambda \ll \lambda_1$.

$$\frac{\Delta\lambda}{\lambda_1} = \frac{v}{c} = \left(\frac{R_2}{R_1} - 1\right) = \frac{R_2 - R_1}{R_1} = \frac{dR}{R}$$

Lemaître assumed that extragalactic nebulae existed at a distance $r$, very close to us, as compared to the radius of the universe, $r << R$. For this case, the spatial part $R(t)d\sigma$ of the line element $ds^2 = -R(t)^2 d\sigma^2 + dt^2$ can be replaced by $r$, so $ds^2 = -r^2 + dt^2$. Furthermore, in this situation, it is valid to set $ds = 0$, since light travels on null geodesics. Therefore, r = $dt$ expresses the time in units of r/c. Hence, Lemaître obtained the following approximate formulae:

$$\frac{v}{c} = \frac{d\mathrm{R}}{\mathrm{R}} = \frac{\dot{\mathrm{R}}}{\mathrm{R}} dt = \frac{\dot{\mathrm{R}}}{\mathrm{R}} r$$

Thus, he obtained the Hubble relation $v = \frac{\dot{\mathrm{R}}}{\mathrm{R}} c \cdot r$ (Equation (24) in Lemaître's paper), which is written as follows in modern terms:

$$v = H_0 r$$

Afterward, using the radial velocities of spirals determined by Strömberg [67] in 1925 and their apparent magnitudes measured by Hubble [68] in 1926, Lemaître calculated the value $H_0$ (known as the Hubble constant today) to be about 625 km/s/Mpc.

In the English translation of Lemaître's paper, produced upon request from Eddington and published in 1931, the brief analysis of the linear velocity–distance relation as well as equation No. 24 (Hubble law), published in the original French version, were omitted in the translated paper [69]. For a long time, the reason for this remained a mystery. Many commentators formulated hypotheses on this matter, ranging from conspiracy to carelessness. But it turned out that it was Lemaître himself who decided not to incorporate them. For Lemaître's, motives an ample explanation is given by Mario Livio's account and the references therein [70].

## 11. The Earliest $H_0$-Tension

Towards the beginning of the 1930s, the main objection to the interpretation of the Hubble constant as the indicator of the universe's expansion was that its measured value implied a universe younger than Earth's accepted age.

Before the 1930s, scientific estimates of the age of the Earth dated from the mid-19th century, from Helmholtz–Kelvin gravitational contraction ages [71] to those resting on geology (i.e., the amount of salt in the oceans, sediments). Finally, after a fierce debate between geologists and physicists [72], an estimate of age based on early-20th radioactive dating was determined [73]. This produced an estimate of around 2 to 3 billion years by the 1930s [74].

On the other hand, the cosmological age of the universe was estimated by taking the reciprocal of the value of the Hubble constant (Hubble's time). Hubble's first evaluation of his eponymous constant resulted in a value of around 500 km per second per megaparsec. This implied an age of the universe of about two billion years, which was in a tense contradiction with the estimated age of the Earth—about three billion years. As a consequence, this mismatch created room for doubt. Some critics questioned that the observed nebulae redshifts were, in fact, a manifestation of Doppler's effect. Such was the paradigm that reigned in the early 1930s.

This incongruity raised two possible explanations: on the one hand, measurements were wrong, or on the other hand, Doppler's shift needed a novel interpretation, perhaps through "new" physics.

Regarding the latter case, in 1929, Fritz Zwicky proposed the concept of "tired light" as an alternative explanation for the redshift–distance relationship [75]. This was a hypothetical redshift mechanism where photons lost energy over time via collisions with other particles in their trajectories through a static universe. Thus, the more distant objects would appear redder than more nearby ones.

Zwicky's idea was not taken lightly by the scientific community, in view of the tension between the apparent age of Earth and the predicted age of the universe. In fact, in 1935, Edwin Hubble and Richard Tolman wrote [76]:

> "... both incline to the opinion, however, that if the red-shift is not due to recessional motion, its explanation will probably involve some quite new physical principles [... and] use of a static Einstein model of the Universe, combined with the assumption that the photons emitted by a nebula lose energy on their journey to the observer by some unknown effect, which is linear with distance, and which leads to a decrease in frequency, without appreciable transverse deflection".

Let us now turn our attention to the possibility of an erroneous Hubble constant measurement in the 1930s. This, as we all know, requires only the measurement of the distance and velocity of a sky object belonging to the galaxy in question. The measurement of recession velocity was straightforward even in those days. Spectral recordings had already achieved good accuracy. However, distances were measured, as we shall see next, by observing standard candles.

## 12. Cepheids

We must now travel back some years to narrate how early distances to nearby galaxies were obtained. Their assessment was necessary in order to achieve a reliable figure for the Hubble constant. It was known from the very beginning that parallax was of little use for faraway objects. Hence, making a comparison between a standard candle and the stellar object whose distance was in question was the next reasonable step to determine its proximity. Cepheid stars were the natural choice.

Classical Cepheids are high-luminosity, radially-pulsating, variable stars. Their intrinsic brightness periodically fluctuates within a range from $-2 > M > -7$ mag at visual wavelengths, making them, in principle, ideal standard candles for distance indicators on galactic and extragalactic scales. The periodic variation of one of these stars was first discovered in 1784 by John Goodricke. He discerned the period of the star δ Cepheid, the prototypical example of a Cepheid [77]. Then, in 1913, as we have already mentioned, the discovery by Leavitt of a P-L (period–luminosity) relation of classical Cepheids residing in the SMC led Hertzsprung to perform a preliminary P-L calibration, but in his work, he ignored interstellar absorption. Afterwards, as we have pointed out, in a subsequent publication following the suggestion made by Edward Skinner King, Russell and Shapley estimated that interstellar absorption diminished stellar magnitudes by about 2 units per kpc.

By 1918, Harlow Shapley had applied a more sophisticated color correction method to the conversion of Leavitt's photographic magnitudes into visual ones, producing what seemed at that time to be a consistent calibration of the P-L Cepheids. However, today, we know that Shapley's calibration was wrong from its beginning. It had a zero point error of around 1.5 in magnitude. The main two reasons for this gross imprecision were that Shapley did not take into account interstellar absorption in spite of his previous work with Russell, and that he unknowingly included the two types of "Cepheids" that we know to exist into today a single set of variable stars. This latter fact was unknown to Shapley.

Let us describe this in more detail. There are two classes of Cepheids: Type I, also known as classical Cepheids, and Type II, sometimes rarely called W Virginis stars after their prototype, W Virginis. Figure 4 shows a modern sketch of regions of type I and type II populations, and the variables RR Lyrae are expected to occur in a P-L graph.

With regard to the RR Lyrae shown in Figure 4, these stars pulse in a manner similar to Cepheid variables. They are commonly found in globular clusters, but the nature of these stars is rather different. It is pertinent to point out that RR Lyraes are also used as standard candles.

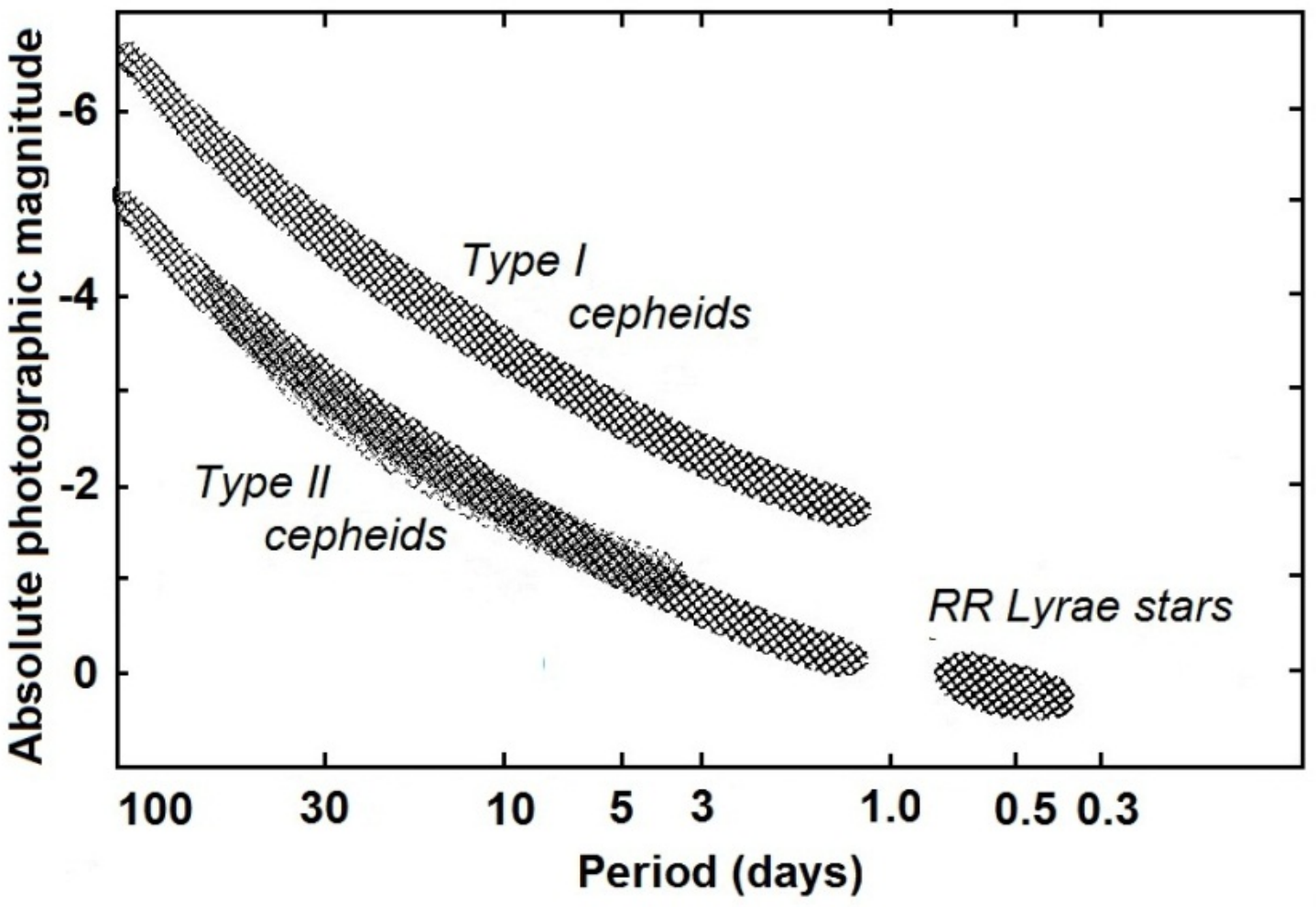

**Figure 4.** Approximate regions of type I, type II, and variable RR Lyrae populations that are expected to appear.

Shapley's way of obtaining a P–L relation was to first inspect data on type I Cepheid stars located in the SMC to determine the zero point of the calibration. For the already-exposed reasons involving absorption, the observed luminosities of these type I stars were dimmer by approximately 1.5 magnitude. Coincidentally, this magnitude difference is what roughly separates type I from type II Cepheid curves (see Figure 4). If the observed type II stars happen to be close to us (i.e., in Milky Way globular clusters), their luminosities are comparatively unaffected by absorption. This means that a type I region apparently overlaps with a type II region (see Figure 5). Another remarkable coincidence is that the slopes of types I and II are very similar to each other.

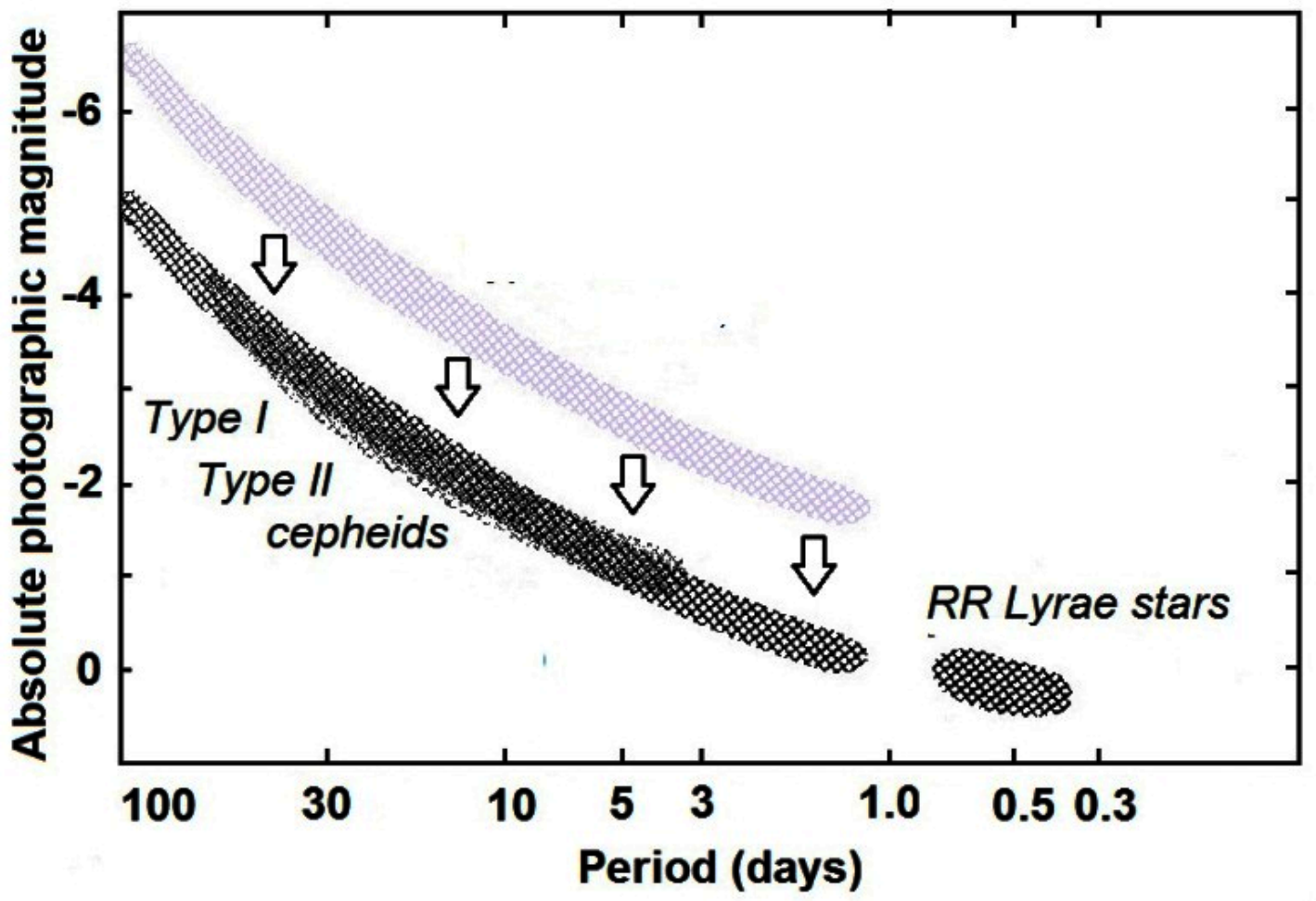

**Figure 5.** Apparent overlap between regions of type I and type II Cepheids. Arrows indicate that type I were misidentified as type II Cepheids. Also shown are RR Lyrae.

In constructing a P-L relation, Shapley not only unknowingly included Cepheids of both types I and II from seven different stellar systems into a single P-L graph, but went even further and also added in a set of RR Lyrae variables. Figure 6 shows the P-L graph that Shapley obtained. There, one can observe that the linear PL relation broke down at the bottom right corner of the figure due to the inclusion of what Shapley called "cluster type Cepheids", now termed RR Lyrae variables. He explicitly made the distinction. The reason for this bending in the P-L linear relation was that, unlike Cepheid variables, RR Lyrae variables do not follow a strict period–luminosity linear relationship at visual wavelengths, although they do in the infrared K band [78].

By the mid-1920s, the Shapley calibration of the P-L Cepheids with its later adjustments was considered the most appropriate and reliable method for determining stellar distances [79,80]. In spite of several justified criticisms (on this matter, see Fernie's review paper [81]), generalized acceptance of Shapley's P-L calibration prevailed for several decades. It is interesting to recall that, as late as 1935, Hubble, in his Silliman Memorial Lectures at Yale, declared regarding Shapley's calibration: "Further revision is expected to be of minor importance" [82].

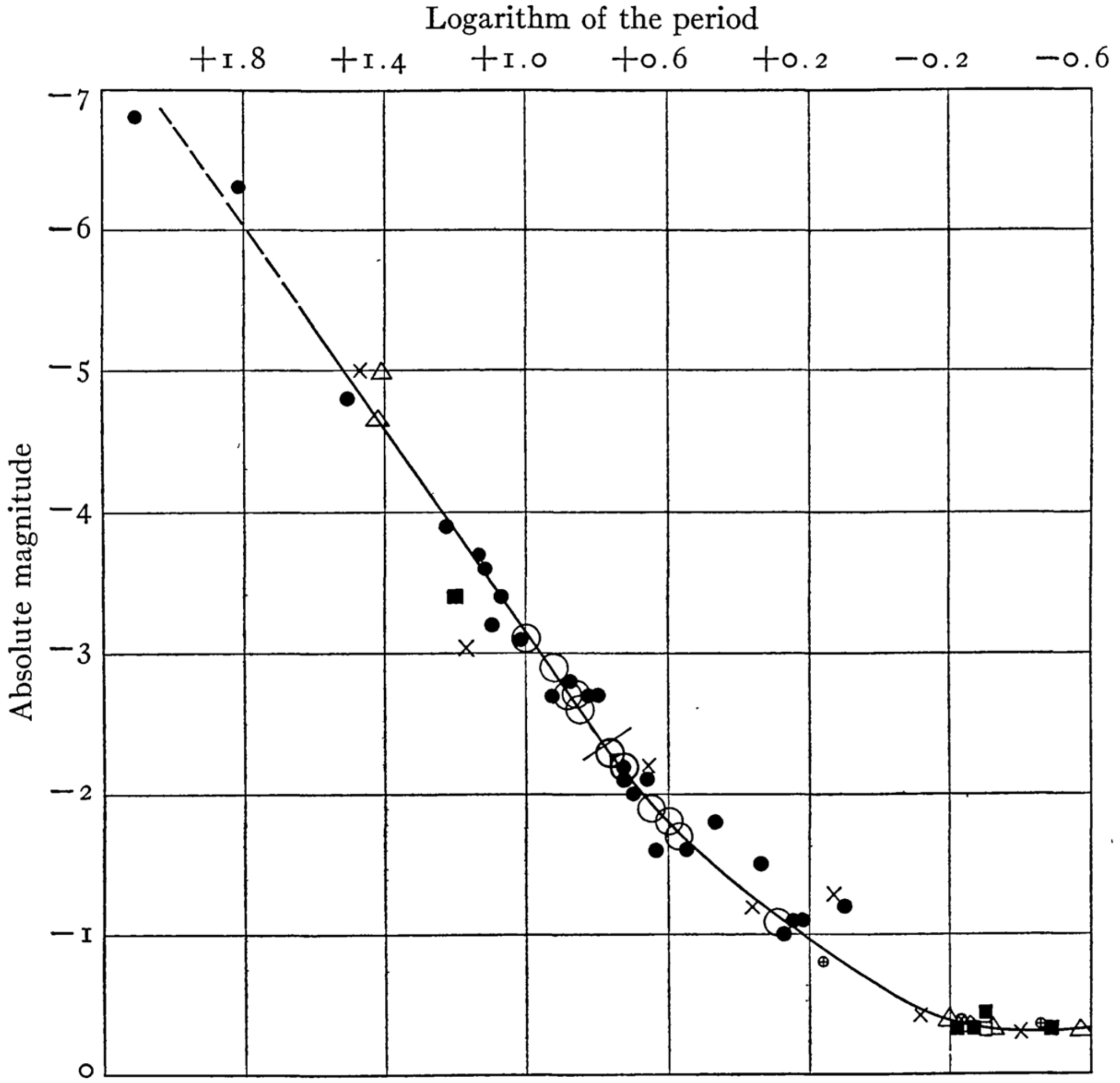

**Figure 6.** Luminosity–period curve of Cepheid variation. The various symbols designate variables from seven different systems. Credit: figure taken from Shapley's 1918 paper [79].

Nevertheless, the extraordinary coincidences that led to Shapley's calibration affected future estimates of Hubble's constant. These coincidences were not spotted for a while, until Walter Baade and Thackeray discovered Shapley's error.

## 13. Baade and Thackeray, a Good Fresh Breeze

In the late 1940s, on cloudy winter nights that hampered observation at Mount Wilson, Edwin Hubble and Walter Baade filled the time by chatting. One of those nights, Hubble mentioned to Baade a concern that had haunted him since 1931, when he inves-tigated the globular clusters of the Andromeda nebula. He noted that the upper limit of luminosity of the globular clusters was about 1.5 magnitudes fainter than the upper limit of those of our own galaxy. Baade argued that this discrepancy was difficult to understand unless "*the discrepancy had entered through some loophole in one of the distance determinations*" [83]. Hubble, in turn, argued that there might be a real difference between the clusters of Andromeda and those of our own galaxy. To Baade's surprise, Hubble added that he found that the brightest globular clusters in M33 were still fainter than those of Andromeda. According to Baade, the discussion ended when they realized that none had a convincing explanation for the discrepancy. This conversation may have seeded doubt in Baade regarding Shapley's calibration, but it was only after the recognition that there were two types of Cepheids that "*the first serious doubts arose concerning the accepted form of the period-luminosity relation*" [83]. The question was, then, how to disprove Shapley's P-L relation.

Fortunately for Baade, the 200-inch Hale telescope at Mount Palomar Observatory was close to completion. This offered him the ideal opportunity to select a nearby galaxy which contained stellar populations of variables so their luminosities could be compared "*side by side*". For Baade, there was no doubt that the Andromeda was the most suitable object for such an investigation and that the 200-inch telescope could answer his questions. However, although the telescope was dedicated in 1948, small optical defects were found, so it took the entire next year of 1949 to correct them. Thus, the telescope was not operational until 1950. Naturally, Baade was very eager to settle these disturbing questions.

In 1949, Baade received a letter dated February 16th from Andrew David Thackeray, based at the Radcliffe Observatory in South Africa, asking for advice. Thackeray explained "*...I have examined several* [globular clusters] *in the Mag.* [Magellanic] *Clouds but I fancy such work will overlap a good deal with Harvard* [Shapley's adscription]...". On March 29th, Baade replied, celebrating that the Radcliffe telescope was in operation and saying that he was sure it would settle many questions that could not be answered from the northern hemisphere. He suggested to Thackeray that, among other topics, it would be highly desirable to investigate whether there were "truly" globular clusters in the LMC. Then, he clarified why he made this odd remark. He explained that Shapley had always been talking about globular clusters in the LMC, but had not yet found variables of the type Baade called globular variables, which were supposed to be there. This seemed very odd to Baade. He further explained that these stars must have existed there, unless the very unlikely fact was true that globular clusters in the Magellanic clouds were of a different nature to those in our galaxy. Recall that this query arose between Hubble and Baade during their chats on cloudy nights at the Mount Wilson Observatory.

Baade ended his letter to Thackeray, not without showing a hint of personal rivalry between him and Harlow Shapley:

> "*...Whatever the final outcome we would know where we stand in the view of a most vexing question. Both Hubble and I hope that Shapley's tendency to consider the Magellanic Clouds as his personal property will not deter you from attacking this problem. He has monopolized the Clouds all too long and it is high time that the barbed wire fences and the warning signs "Keep out. This means you!" are taken down. Monopolies in science are intolerable and should never be respected. Moreover lately Shapley has worked his gold mine only if he needed money for booze (some stuff for publication). The whole situation has become intolerable and a good fresh breeze is most desirable...*" [84].

For his part, the research program undertaken by Baade using the powerful and brand-new Palomar telescope was intended to identify variable stars and compare their luminosities, since he suspected that, in reality, the origin of the problem with Shapley's calibration was not the existence of different globular clusters, but of two different populations of Cepheids. Three years later, in 1953, at the IAU General Assembly in Rome, Walter Baade reported his findings [85].

Baade reported that RR Lyrae variables could not be detected in the Andromeda Nebula, even when using the 200-inch telescope. The reason for this negative result, according to Baade, was that Shapley's calibration underestimated distances. His reasoning was simple and straightforward: Baade assumed that if the Shapley calibration was correct, the distance to Andromeda would be around 275 to 300 kpc, and in that case, it would be simple and straightforward to compare the variables' luminosities with the use of the 200-inch telescope. Instead, he could only observe the very brightest Population II stars on limiting exposures, but not a single RR Lyrae star. Baade concluded that Shapley's calibration underestimated the distance to M31, and that RR Lyrae stars were not visible because they exist much further away. The argument made by Baade seemed convincing, but strictly speaking, it was not an incontrovertible reason to assert the failure of Shapley's relation, only indicating a very strong possibility.

However, the incontrovertible evidence came immediately after Baade had spoken, when Thackeray announced that he and A.J. Wasselink had discovered the first Lyrae variables in the SMC globular cluster (NCG 121). Recall that the SMC was the galaxy in which Leavitt originally established the first P-L relation using classical Cepheids.

This discovery meant that the absolute magnitudes of RR Lyrae and Cepheids (type i) could be directly compared. According to Shapley's calibration, these RR Lyrae should have appeared at a magnitude of 17.5, but it should be of no surprise to the reader that they were actually fainter by 1.5 mag. This proved Shapley's mistake.

As result, the distance to M31 was twice as far as originally calculated. The perceived universe doubled in size; the Hubble constant reduced its value by half; and the age of the universe doubled, partially easing the paradox of Earth being older than the universe.

In retrospect, Baade commented on this years later: "…there were good reasons to suspect that unknowingly Shapley had made a fatal step when he linked the cluster-type variables to the type I Cepheids through the type II Cepheids in globular clusters and that in reality we were dealing with two different period-luminosity relations, the one valid for the type I Cepheids, the other for the type II Cepheids".

### 14. The Changing Value of $H_0$

In this section, we do not attempt to list all measurements made during the period from Lemaître's first estimate of $H_0$ in 1927 to the mid-twentieth century measurements. This has already been carried out many times, often by the original researchers or their close collaborators (see, e.g., Fernie 1969 [81]). Our intention here is to show the downward trend experienced over time during the second quarter of the last century for the values of the Hubble constant. This trend was not due to an arbitrary value-adjustment of $H_0$ from the observers, but to efforts to improve observations, "revise" bias and confusions made by others on observed objects, and to consider the sources of errors that were previously overlooked or ignored (shown in Table 1). As we have already seen in the previous section, by the mid-twentieth century, Shapley's calibration had been found to be erroneous and was substituted by Baade and Thackeray's calibration.

| Year | Author | Value (km/s/Mpc) | Method | Ref. |
|---|---|---|---|---|
| 1927 | Lemaître | 600 | Used data of Hubble and Strömberg | [86] |
| 1929 | Hubble | 500 | Cepheids; Shapley's calibration P−L | [38] |
| 1931 | Hubble and Humason | 526 ± 10% | Cepheids; Shapley's calibration P−L | [39] |
| 1946 | Mineur | 320 | Interstellar absorption corrections making a major recalibration to Shapley's scale. Used zero point $M_0 = -1.54$, see Baade (1956) | [87, 88] [83] |
| 1951 | Behr | 240 | He heuristically pre-discovered the Scott effect and made the co-rresponding corrections to obser-vationsby J. Stebbins and A. E. Whitford (Astrophys J. 108 413 (1948), Mt Wilson Contr. 753, using Shapley's calibration | [89] |
| 1952 | Baade and Thackeray | 280 ± 30 | RR Lyrae stars; corrected Sha-pley's calibration (see next sec-tion) | [84, 90] |
| 1956 | Humason, Mayall, and Sandage | 180 ± 20 | Comparison of photographic magnitudes for 576 galaxies to their redshifts; their calibration set the brightest galaxies in clusters equal to M31 luminosity | [91] |

Inherently, the observed trending decrease in the value of the Hubble constant increased the Hubble time. This, in turn, reduced the tension that existed between the supposed age of the universe and its estimates obtained by various means, namely, from estimates made according to the ages of some stars. Figure 7 shows some selected Hubble constant values in relation to the decreasing trend that $H_0$ estimates experienced during the second quarter of the twentieth century.

**Table 1.** Trend of values for $H_0$ in the period of 1930–1955.

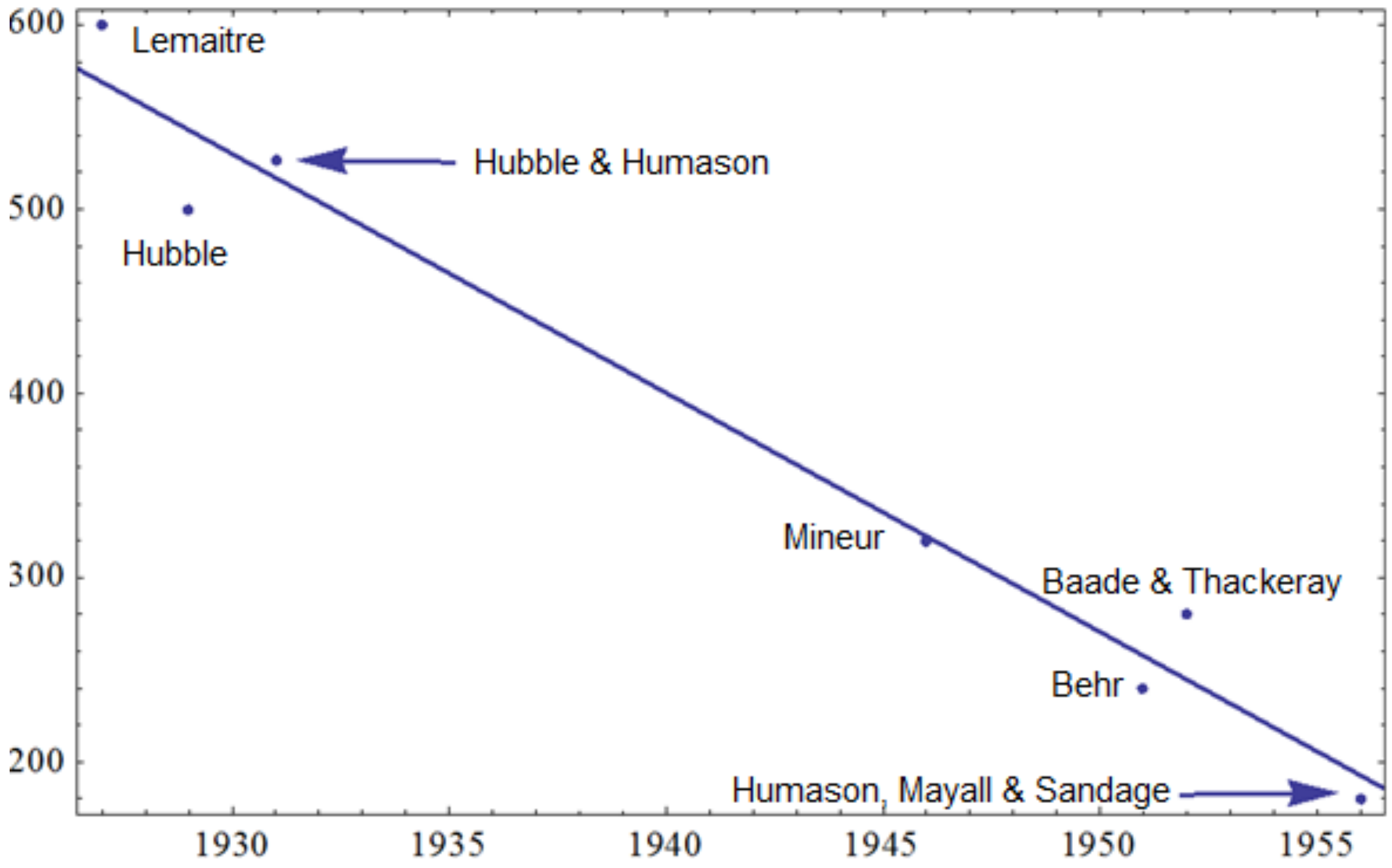

**Figure 7.** Some values determined for $H_0$ in the period of 1930–1955, showing the trend.

## 15. The Need to See Further

When Walter Baade and David Thackeray revealed that Hubble had made a miscalculation of the distances to galaxies by using a wrong Cepheid gauge calibration, they erased, at a stroke, the issue on the supposed young age of the universe—it was confirmed to be older.

However, a hurdle would soon arise on the use of Cepheid stars to determine the value of $H_0$. This obstacle was that, even though the Cepheid stars were bright (on average, about $10^4$ brighter than the sun) their luminosity was not sufficient for them to be observed them at that time beyond a few dozen parsecs. On this scale, this meant that the universe could not be assumed to be isotropic or homogeneous.

In addition, it had been known since Vesto Slipher's early measurements that not all nearby galaxies were receding; on the contrary, a pair of them was actually approaching us. This meant that, although nearby galaxies participate in the expansion of the universe, gravitational interactions among neighboring galaxies may cause some of them to move much faster or slower than the rest of the universe. In fact, it was later discovered that our local group of galaxies experiences a "Virgocentric flow". This is a preferential movement directed towards the Virgo Cluster, so corrections must be applied to obtain the Hubble flow (see Figure 8).

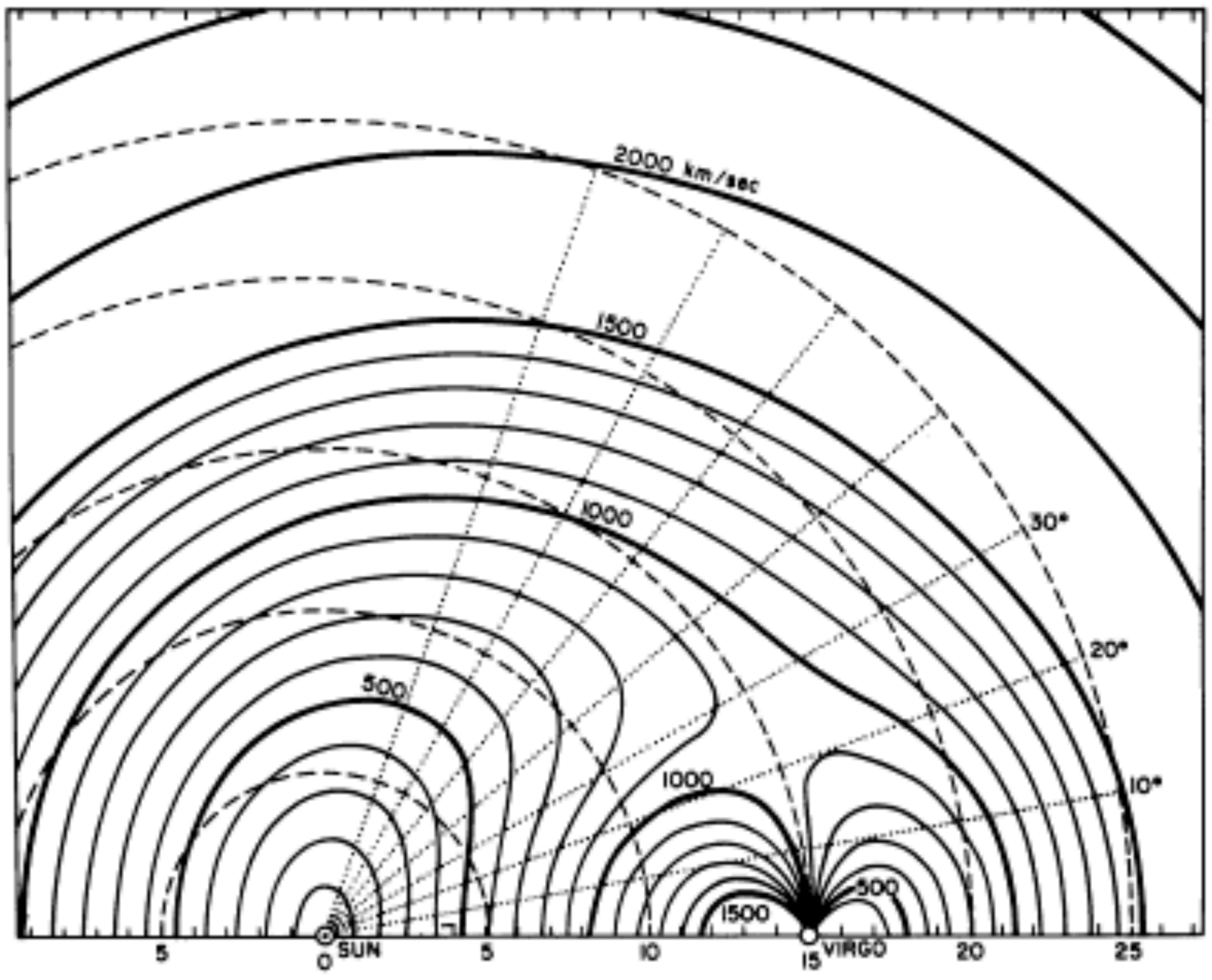

**Figure 8.** One of two 2D grids with Earth and the Virgo cluster on the *x*-axis. Redshift contours are plotted for a Virgocentric flow. Note that a pure Hubble flow would be concentric (from Tonry and Davis 1981) [92].

The fact that the genuine Hubble flow is affected by local gravitational interactions forced astronomers to accurately determine the Hubble constant at remote distances. The need to measure further away was clear, as the large-scale structure of the universe makes it isotropic. This task is extremely difficult.

**Part II. Building and adding rungs to the cosmic ladder**

During the second half of the last century, several independent groups of astronomers developed methods for measuring the distances to galaxies, leading to new estimates of the Hubble constant and generating debates on the merits of each technique. We shall next provide a brief description of the most relevant methods that appeared in the second half of the 20th century.

In 1970, Sidney Van den Bergh [93] from the David Dunlap Observatory in Toronto reviewed nine methods used at that time to determine the Hubble constant. These were: diameters of HII regions (obtaining $91^{+19}_{-15}$ km/s/Mpc), luminosity classification of galaxies ($98^{+19}_{-17}$ km/s/Mpc), brightest globular clusters in galaxies ($72^{+15}_{-12}$ km/s/Mpc), mass-to-light ratios ($105^{+33}_{-26}$ km/s/Mpc), third-brightest cluster galaxy ($126^{+48}_{-35}$ km/s/Mpc), supernovae and the extra-galactic distance scale ($123^{+47}_{-34}$ km/s/Mpc), surface brightness and diameter of galaxies ($89^{+46}_{-30}$ km/s/Mpc), brightest stars in galaxies ($95^{+15}_{-12}$ km/s/Mpc), and regional variations of the Hubble constant (not specified). Van den Bergh computed their mean value, obtaining $95^{+15}_{-12}$ km/s/Mpc.

These methods, as originally thought, are no longer employed today. New methods were to arise in the following years. We account for them in the following sections.

## 16. Planetary-Nebula Luminosity Functions

In simple terms, planetary nebulae (PNe) are expanding shells of bright, luminous gas expelled by dying stars of low-to-intermediate mass (~1 to 8 $M_\odot$) towards the end of their evolution. Planetary nebulae is an inaccurate term for these objects because they are unrelated to planets or exoplanets. PNe are important tools used to estimate distances, despite the brief lifespans of their terminal phases (~30,000–50,000 y).

In 1963, Karl G. Henize and Bengt E. Westerlund suggested that PNe in the Milky Way and in the SMC might all have the same maximum brightness [94]. Their hypothesis opened the way to using bright PNe as extragalactic standard candles. But it was not until 1978 that Holland Cole Ford and David Charles Jenner estimated the distance to the large spiral M81 by making observations on eight PNe in the exterior of this Local Group galaxy [95]. They noted that these PNe appeared to be around 20 times fainter than those found in M31. Hence, assuming that the suggestion made by Henize and Westerlund of maximum brightness equality was valid, they estimated that M81 was about 3 Mpc away. This distance agreed quite well with other independent assessments made at that time.

The advent of charged coupled devices (CCDs) in the early 1980s, combined with superior telescopes, allowed Ford, together with George Howard Jacoby and Robin Bruce Ciardullo, to develop an alternative and superior method by which to use PNe as distance indicators [96–99]. They employed the concept of luminosity function, which characterizes the distributions of PNe by counting the total number of PNe in each luminosity or absolute magnitude interval in a given volume. Jacoby and Ciardullo used a narrow band filter to register the [O III] emission-line (λ5007) of PNe [100]. These objects glow predominantly at this wavelength, so this facilitated the subtraction of sky backgrounds. In this way, Jacoby and Ciardullo found and recorded several hundred extragalactic PNe.

During their investigation, Jacoby and Ciardullo noted that, when they studied a galaxy in this way, they detected many faint PNe, but few bright PNe. By counting the number of PNe of each level of brightness, they defined the planetary nebulae luminosity function (PNLF) [101]. The important finding was that the shape of the PNLF appeared to be the same for every galaxy.

The number of PNe decreased smoothly with the increasing luminosity, and there was a limiting luminosity beyond which they could not detect more PNe. The cut-off was sharp. Jacoby and colleagues proposed that the PNLF was of the form suggested by Ciardullo et al. (1989) [98]:

$$N(M) \propto e^{0.307M}\left(1 - e^{3(M^* - M)}\right),$$

where *N(M)* is the number of PNe with absolute magnitude *M*, and M* is the absolute magnitude of the most luminous PNe (i.e., the bright cut-off magnitude), which is currently accepted to be [102] M* = 4.47 ± 0.05. This characteristic cut-off in the PNLF provides an excellent standard candle. A comparison of the PNLF in a remote galaxy with that in M31 gives the distance in terms of the M31 distance. Figure 9 shows the PNLF for M31.

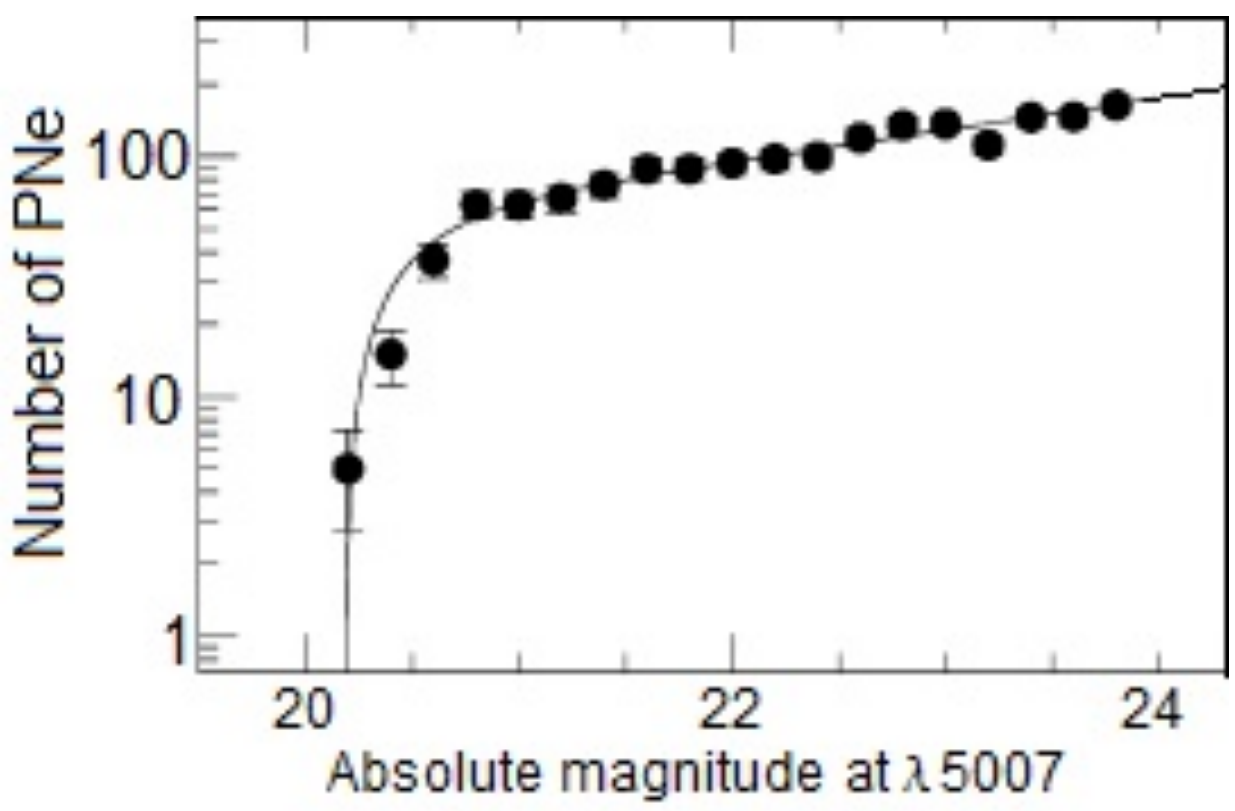

**Figure 9.** The planetary luminosity function for M31 from Merrett et al. [103]. The sharp cut-off at the bright end of the [O III] PNLF is characteristic of all galaxies. Note the good fit of the standard PNLF.

Before satellite-based telescopes, ground-based observations using the PNLF were limited to about 20 Mpc. Beyond that distance, dimmer PNe become too faint to observe. Therefore, it was a useful complement to calibrations using Cepheids. Today, the cut-off of PNLF constitutes a robust standard candle. It is now used to calibrate Type Ia supernovae (e.g., Jacoby et al. 2006 [104]).

One the questions concerning the [OIII] PNLF technique is why there is such observational invariance in the bright cut-off magnitude along the morphological sequences of galaxies, i.e., from spirals undergoing active star formation to the old populations in elliptical and lenticular galaxies. This is a current point at issue, and the answer has not yet been firmly established. However, the PNLF helps to tie together Population I and Population II strands in the distance ladder [105].

## 17. The Tully–Fisher Relation (TFR)

In 1977, Richard Brent Tully and James Richard Fisher discovered an empirical power law relation between the luminosity of a spiral galaxy and its rotational velocity [106]. Their proposal was to measure the HI 21 cm line width to find the rotational velocity of the galaxy; use this to calculate the luminosity; and compare it with the apparent magnitude of the galaxy to deduce its distance.

To check their proposal, they observed spiral galaxies in two nearby clusters, namely, the Virgo cluster and the Ursa Major cluster. Then, they made a comparison between galaxies in the same cluster. This assured them that, in principle, galaxies are situated essentially at the same distance. Finally, by applying their idea to nearby, Cepheid-calibrated galaxies, Tully and Fisher judged their method, and it agreed with previous distance measurements. The numerical value they obtained for $H_0$ using TRF in 1977 was 80 km/s/Mpc.

Subsequent work by Marc Aaronson, John Huchra, and Jeremy Mould [107] found that the power law relationship favored a power law exponent to the fourth, $L \approx V^4$, if the spirals possessed some properties. In 1983, Aaronson suggested an improvement to the TFR method by measuring luminosities in the infrared (K band) radiation rather than the optical band to reduce scattering effects [108].

Since then, the TFR started serving as a rung on the cosmic distance ladder. TFR was calibrated by means of the Cepheid gauge. By the end of the 20th century, TRF had established itself as standard technique, extending the ladder to a greater distance.

By the turn of the century, it was clear that the TFR fit better when luminosity was replaced by the galaxy's total baryonic mass [109]. This latter variation is known as the baryonic Tully–Fisher relation (BTFR). It states that baryonic mass is proportional to velocity, again, to the power of roughly 4 [110].

## 18. The Faber–Jackson Relation (F-J)

During the 1970s, efforts were carried out to find possible correlations between various parameters of elliptical galaxies, such as their luminosities, sizes, central velocity dispersions, abundances of heavy elements, etc.

In 1976, the astronomers Sandra Moore Faber and Robert Earl Jackson discovered a relationship connecting the luminosity of elliptical galaxies to the velocity dispersion (σ) of stars near their respective centers [111]. They found that the luminosity of an elliptical galaxy increases with the velocity dispersion of the stars to a power of n ($L = \sigma^n$). Originally, Faber and Jackson found that $n \approx 4$. But later, it was found that the value of n depends on the range of galaxy luminosity. For low-luminosity elliptical galaxies, the F-J relation is fitted with a value of $n \approx 2$. This value was found by a team led by Roger Davies [112]. A value of $n \approx 5$ for luminous elliptical galaxies was reported by Paul L. Schechter [113]. These findings have been confirmed observationally by many authors. For a comprehensive reference list in the literature, see, for instance, Markovic and Guzman [114].

Like the TFR, the F-J relation has been useful in providing another means of estimating distances, in this case to elliptical galaxies. One possible advantage of the F-J method is that it uses the more easily observable properties of elliptical galaxies by measuring the central stellar velocity dispersion, which can be achieved relatively easily by using spectroscopy to measure the Doppler shift of light emitted by the stars. Afterwards, an estimate of the true luminosity of the galaxy via the F-J relation can be obtained. Finally, a comparison can be made to the apparent magnitude of the galaxy. This comparison provides an estimate of the distance modulus and, thus, the distance to the elliptical galaxy. However, there is still a major disadvantage. As we have already pointed out, elliptical galaxies do not faithfully follow the F-J power law, since their coefficient n varies from 2 to 5 depending on their luminosity.

In 1991, Donald H. Gudehus suggested combining a galaxy's central velocity dispersion with measurements of its central surface brightness and radius parameter [115]. By using this combination, or "reduced galaxian radius parameter", Gudehus asserted that it was possible to improve the estimate of the galaxy's distance, yielding results free of systematic bias.

## 19. Fundamental Plane, the $D_n$-$\sigma$ Relation

During the decade that followed the discovery of the F-J relation, extensive research was carried out attempting to find out other possible correlations between various parameters in elliptical galaxies. But it was not until 1987 that the American astronomers Marc Davis and Stanislav George Djorgovski [116] replaced the luminosity of an elliptical galaxy in the F-J relation with two novel parameters: the effective radius of the elliptical galaxy ($R_e$) and the average surface brightness ($I_e$) within that radius. They found a relationship connecting the three parameters: $R_e = k\, \sigma^{1.36} I_e^{-0.85}$, where $k$ is a constant.

Any one of the three parameters (average surface brightness, velocity dispersion, and effective radius) may be estimated from the other two; together, they describe a plane which Davis and Djorgovski called "the fundamental plane" for ellipticals. In the Davis and Djorgovski method, $I_e$ and σ are measured to obtain an effective radius $R_e$. This effective radius is used as a standard rod. Once this rod's length has been obtained, by directly measuring its angular size, it becomes easy to determine the distance from the observer to the galaxy through small-angle approximation.

The Davis and Djorgovski method stands out against the F-J relation, where σ is measured and luminosity is used as a standard candle. Incidentally, the F-J relation can be viewed as a line projected on the fundamental planes of elliptical galaxies.

An alternative variant correlation in the fundamental plane was proposed in 1987 by the group of the "Seven Samurais", as they were known (1987) [117]. The group was composed of American, Argentinian, and British astronomers: David Burstein, Roger Davies, Alan Dressler, Sandra Faber, Donald Lynden-Bell, Roberto Terlevich, and Gary A. Wegner. The Seven Samurais found that there is an excellent correlation between σ and a quantity they called $D_n$, which represents the diameter of a central circular region of an elliptical galaxy within which the total average surface brightness is some particu-lar value (20.75 magnitude per square second of arc). The correlation can be expressed as:

$$\frac{D_n}{kpc} = \left( \frac{\sigma}{100\,\mathrm{km/s}} \right)^{1.33}$$

This relationship has a scatter of 15% between galaxies, as it represents a slightly oblique projection of the fundamental plane. The "Seven Samurai" published a series of papers providing the results of an all-sky survey of an elliptical, building a large store of data on elliptic galaxies.

## 20. Surface Brightness Fluctuation Method (SBF)

The surface brightness fluctuation method (SBF) was proposed in 1988 by the American astronomers John Landis Tonry and Donald P. Schneider [118], and later was further developed by Tonry [119].

This method aims to estimate the distance from an observer to a galaxy by measuring the degree of brightness fluctuations at its surface. The most common situation for a telescope is not being able to resolve a population of bright stars in distant galaxies. However, the discrete nature of stars causes fluctuations in the surface brightness of the galaxy. The further away the galaxy is located, the more "diffuse" and less "dotted" it will appear.

The SBF method measures the variance in a galaxy's light distribution, which arises from fluctuations in the amounts and luminosities of individual stars for each resolution element.

The galaxy surface brightness is independent of the distance from the galaxy to the observer, but the variance (measured in the Fourier space) is not. This variance changes as the inverse of the square of the distance "d" in which the galaxy is located ($d^{-2}$).

In practice, the CCD image of the galaxy under observation must be excised from external background and foreground sources. Then, an isophotal model is fitted to the image. Next, the model is subtracted from the galaxy image. The remaining image is Fourier-transformed, and the SBF pattern is measured according to the power spectrum of the residual image that is left behind. The amplitude of the spectrum gives the luminosity of the galaxy. Pertinent corrections for interstellar dust absorption must also be accounted for. Tonry and colleagues measured the Hubble constant for over a decade, obtaining a range of values from 88 to 77 km/s/Mpc [120,121].

## 21. Tip of the Red Giant Branch (TRGB)

The foundation of this method is similar to that of the previous one, which is based on the constancy of luminosities of stars at the "tip" of the red giant branch in the Hertzsprung–Russell diagram. This happens when the degenerate helium nuclei in low-mass stars reach their critical mass and start burning carbon. Thus, the star is in the horizontal branch afterwards. The TRGB luminosity is typically measured in the I-band, where the luminosity has little dependency on stellar age or stellar metallicity, so it is a well-established standard candle. With the advent of CCDs, this method was consolidated [122–125].

By the mid-1990s, the most reliable primary distance indicator for nearby galaxies was the Cepheid P-L relation, which served as a basis for other secondary distance indicators of other, more distant targets. During this time, the TRGB was shown to be an independent distance calibrator which, in fact, was applicable to a wider type of galaxies, not exclusively to the late-type galaxies used for Cepheid calibration. Both methods were then competing in terms of their precision in determining distances [126]. At that time, in a Space Telescope Science Institute meeting on the extragalactic distance scale, Allan Sandage reported a value of $55^{+5}_{-5}$ km/s/Mpc, while the TRGB group reported $73^{+6}_{-6}$ (statistical) $^{+8}_{-8}$ (systematic) km/s/Mpc [127]. As Wendy Freedman put it in a short review in 1998 [128], the results seemed to reach an uncertainty of 10% for the Hubble constant when it was measured using different methods, leaving in the past a factor of two that astronomers had used throughout the previous two decades.

The Carnegie–Chicago Hubble Program (CCHP), put forward later, in the 2010s, was designed to provide an alternative route for the calibration of SNe Ia and, thereby, to yield an independent determination of $H_0$ via measurement of the TRGB in nearby galaxies. Their results will be outlined below. The most recent TRGB result was 72.94 ± 1.98 km/s/Mpc, with an additional uncertainty due to algorithm choices of 0.83 km/s/Mpc [129].

## 22. Global Cluster Luminosity Function (GCLF)

Globular clusters (GC) are compact and spherical sets of stars found in the halos of large galaxies. They are among the most luminous objects in galaxies. Due to their brightness, as early as 1955, they were recognized as potentially important distance indicators. It was in this year that William Alvin Baum tried to use a GC to estimate the distance to a particular galaxy. Baum's strategy involved a simple comparison between the brightest GC in M87 (in the Virgo cluster) to that in M31, assuming that the luminosities were identical. Then, he calculated the distance to M87 with the inverse square law, using M31 as a gauge [130]. Incidentally, in his publication, he deduced from his distance measurement a value of 150 km/s/Mpc for $H_0$.

Later, it was observed by Gérard de Vaucouleurs, and independently by other astronomers, that the brightest star clusters of a galaxy are correlated to the luminosity of the galaxy to which they belong [131]. Thus, the assumption made by Baum in 1955 of a fixed luminosity for the brightest clusters in a galaxy was incorrect.

In the late 1970s, René Racine suggested that the globular cluster luminosity function (GCLF) could be a "potentially important distance indicator in extragalactic studies" [132]. His working hypothesis was that the magnitude distribution of old globular clusters exhibited a universal shape. This distribution was characterized by its mean value and standard deviation. In turn, the mean value was used as the standard candle to evaluate distances. For unknown reasons, the shape of the GCLF appeared to be lognormal in shape, with a universal mean and standard deviation (nonetheless, with some dependences on other factors, such as metallicity).

As first introduced by David Hanes, astronomers usually fit a Gaussian curve to $\varphi(m)$, the relative number of globular clusters, as a function of their magnitude, $m$ [133]:

$$\varphi(m) = A\, e^{-(m - m_0)/2\sigma^2},$$

where $m_0$ is the peak or turnover point TO (i.e., the magnitude at which most GC are found), σ is the width of the distribution, and A is the normalization constant. The distribution is then characterized by only two parameters, σ and TO. The latter is the standard candle, and the distance measurement is relative to the Milky Way GCLF TO value or to that of M31. The GCLF is a secondary distance indicator, since the absolute distances to either our galaxy or M31 globular clusters must be known. This method has been used to estimate distances and, hence, the Hubble constant as well.

## 23. Cepheid Calibration Development in the Twentieth Century

In the early 1990s, a group of experts on cosmic distance determination wrote an article discussing which were the most reliable methods for that purpose [102]. These were according to their opinion: planetary nebula luminosity functions (PNLF) (cf., Section 16), Tully–Fisher relations (cf., Section 17), fundamental-plane relationships for elliptical galaxies (cf., Section 19), surface brightness fluctuations (SBF) (cf., Section 20), Globular-Cluster Luminosity Functions (GCLF) (cf., Section 22), and Type-la supernovae (SN la) (see Section 24). We have already described these in the present work, except for the SN Ia method, which we shall describe in the next section.

Before we report on the SN Ia method, we must note that the methods already described so far are all secondary distance indicators; that is, they need to be calibrated, usually by finding a Cepheid distance to a galaxy which in turn requires knowledge of an accurate period–luminosity relation. Cepheid variable stars are still primary calibrators for secondary standard candles. Because of their central role in establishing the distance scale, we must mention the evolution of Cepheid's period–luminosity relation values during the twentieth century.

As we have already stated, it took some years after the revelation of Baade and Thackeray that the P-L relation was wrong for the astronomical community to initiate the search for a reliable and accurate Cepheid P-L calibration. Over the next decades of the second half of the last century, several independent groups of astronomers made great efforts to improve the Cepheid period–luminosity relation. Table 2 shows some of the most influential calibrations made during the 20th century.

**Table 2.** Leading calibrations of the Cepheid period in the twentieth century. Luminosity relation (<Mv> = -a – b log10 P, P (days)).

| Year | a | b | <Mv> | Author |
|---|---|---|---|---|
| 1913 | 0.60 | 2.10 | −2.70 | Hertzsprung [134] |
| 1918 | 0.72 | 2.10 | −2.82 | Shapley [135] |
| 1961 | 1.67 | 2.54 | −4.21 | Kraft [136] |
| 1968 | 1.43 | 2.80 | −4.23 | Sandage and Tammann [137] |
| 1987 | 1.35 | 2.78 | −4.13 | Feast and Walker [138] |
| 1991 | 1.40 | 2.76 | −4.16 | Madore and Freedman [139] |
| 1997 | 1.38 | 2.77 | −4.15 | Tanvir [140] |
| 1997 | 1.43 | 2.81 | −4.24 | Feast and Catchpole [141] |

It is noteworthy that, in the last decade of the twentieth century, the parameters of the period–luminosity relationship began to concur with each other.

In the late 1960s, Wisniewski and Johnson [142] presented high-precision photometry of Milky Way Cepheids that led to the understanding of two properties of light curves: (1) the decrease in amplitude moving from the ultraviolet to the near-infrared, and (2) the shift in the relative phase of maximum light, also with the increasing wavelength. These represented major advances in our understanding of Cepheids [143].

**24. Type Ia Supernovae (SN Ia)**

Since the initial surveys of intergalactic distances, Edwin Hubble had a feeling that novae, as they were generically called (novae and supernovae were not differentiated at that time), might serve as criteria for determining intergalactic distances. His belief was based on the fact that their great luminosity made them detectable at great distances and that, despite the scant available data, it seemed that their maximum brightnesses reached similar standard values. However, he lamented that the sporadic frequency of their appearances in the sky would prevent them from contributing much to the problem of obtaining distances [144]. Walter Baade and Fritz Zwicky coined the term supernovae in the abstract of their joint paper ``Supernovae and Cosmic Rays”, presented orally by Zwicky at an American Physical Society meeting at Stanford in December of 1933. The term supernova was created to distinguish them from ordinary novae, which are far less luminous [145]. It is interesting to mention that, at the time of the publication of Hubble’s “Realm of Nebulae” (1936), only 115 events (novae and supernovae all together) had been recorded [144].

Beginning in 1937, Rudolph Leo Bernard Minkowski, while working at Mount Wilson, made spectroscopic observations of supernovae. Within a few years, he and Baade realized, based on their spectra, that there were two different types of supernovae: Type I and Type II. The first type shows no hydrogen spectroscopic lines, whereas Type II supernovae have hydrogen. By 1986, S. E. Woosley and T. A. Weaver had evaluated accumulating evidence revealing that Type I SN can be separated in into three subclasses [146]: Type Ia class is silicon-rich, Type Ib is helium-rich, and the objects which have neither silicon nor helium in abundance are classified as Type Ic. In turn, the Type II class is divided into II-P, which have ~100-day "plateaus" in their light curves; II-L, which have "linear" declines in their light curves; and II-n, which have narrow lines in their spectra [147].

Type Ia supernovae (SN Ia) are the most luminous and homogeneous kind. An SN Ia supernova event is the blast that occurs when a dead star becomes a natural thermonuclear bomb. They originate from old, low-mass stars in binary systems. To be precise, a white dwarf star near the Chandrasekhar limit (1.4 $M_\odot$) accretes mass from its companion star until it becomes denser, and a thermonuclear blast occurs. The explosion blows the dwarf star completely apart, spewing out material at remarkably high speeds ($10^4$ km/sec). The glow of this expanding fireball takes a few weeks to reach its peak brightness, and afterwards, its luminosity declines over a period of months in a normal way. Figure 10 shows a typical light curve of an SN Ia.

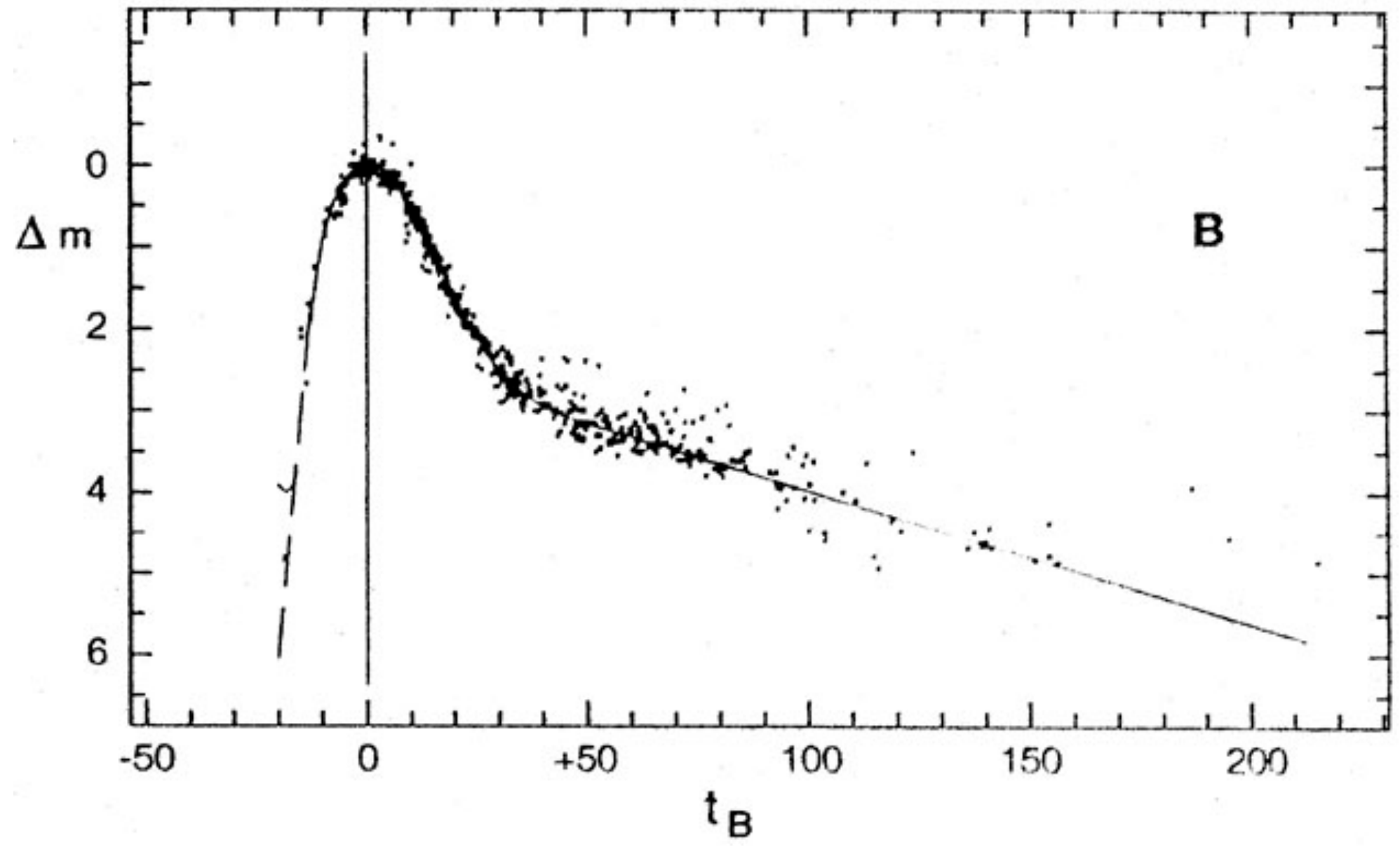

**Figure 10.** Typical B light curve of an SN Ia ($t_B$ = days). Credit: Figure taken from the review cited in [148], based on data of 22 supernovae, by R. Cadonau and B. Leibundgut [149].

In 1968, Charles T. Kowal of Mount Wilson and Palomar published a redshift–magnitude graph relation for 19 type I supernovae, assuming a value for the Hubble constant (100 km/s/Mpc) [150]. Figure 11 shows little dispersion of his plotted points, concluding that "There is therefore considerable hope that the magnitudes of type I supernovae, [...], can be used as reliable distance indicators, [...], and visible at very great distances".

In his 1968 paper, Kowal also predicted the future use of supernovae to determine cosmic acceleration: "It may even be possible to determine the second order term in the redshift-magnitude relation when light curves become available for very distant supernovae". The "second-order term" would be the one that indicated cosmic acceleration or deceleration". This was proven to be the case almost three decades later. It is pertinent to mention here that the original idea of tracking supernovae as deceleration indicators goes back to the 1930s, when Walter Baade and Fritz Zwicky thought it could be possible to learn how much the universe was decelerating by observing supernovae [151].

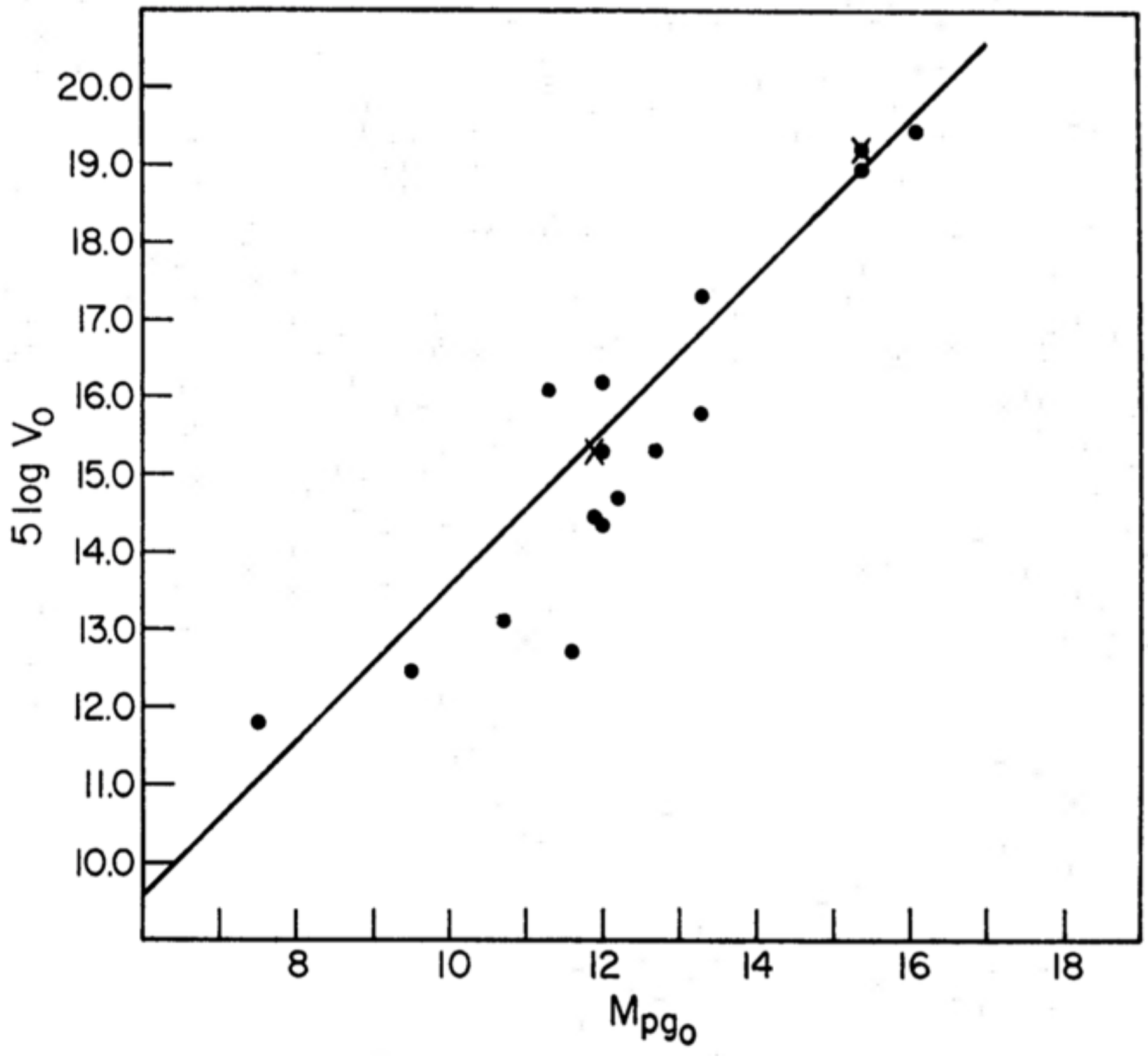

**Figure 11.** Kowal's redshift–magnitude relation. The crosses represent the average velocity and average magnitude of supernovae in the Virgo and Coma clusters (Kowal 1968 [150]).

On 22 June 1990, E. Thouvenot, at the Observatory of the Côte d'Azur, discovered a supernova SN 1990N significantly before its maximum [152]. This timely discovery (14 days before peak luminosity) allowed astronomers at the Cerro Tololo Inter-American Observatory (CTIO) to opportunely record its light curves [153]. Five years later, Allan Sandage asked the Hubble Space Telescope (HST Proposal, 5981) to allocate time for searching Cepheids in the SN 1990N parent galaxy [154]. As result, 20 Cepheids were located. Sandage and collaborators obtained the distance to the parent galaxy by measuring the periods and magnitudes of 20 Cepheid variable stars with HST. With this assessment, they obtained a value of 58 ± 4 km/s/Mpc for the Hubble constant based on a Cepheid–supernova distance scale [155].

Throughout all these years of sporadic observations of supernovae, it became clear that the use of Type Ia supernovae as standard candles had potential to become one of the best methods of determining distances. This is due to their fantastic luminosities and the small dispersion in their inherent peak brightness. However, it was also known that there is a drawback: their visible occurrence is infrequent, and they do not last long at their peak brightness.

This motivated several astronomers to start collaborative ventures using CCD coupled with large telescopes located in different countries. However, this was not an easy task. Before the launch of the HST, a group of Danish astronomers were among the first to use CCDs to search for SN from an Earth-grounded telescope [156]. After a two-year search, in 1985, they managed to find a single SN Ia [157]. It was clear that the job of discovering new SN required a larger collaborative effort, and that they needed to wait for the launch of the HST.

Two years later, in 1987, Saul Perlmutter and Carl Pennypacker, both researchers in Rich Muller's group at UC Berkeley were encouraged by the introduction of new sensitive CCD sensors into astronomy, as well as the new computers that were just then becoming fast enough to analyze significant amounts of data. They decided to measure the deceleration of the universe's expansion using supernovae. The scheme of a universe slowing down its expansion, possibly coming to a halt and, furthermore, perhaps collapsing, was still prevalent at that time.

Based on the Danish group's unfortunately meager results (as we have mentioned above), the approach the Berkeley group took was to develop the capability to gaze at more than a small cluster of galaxies at a time (as the Danish program had achieved). Thus, the Berkeley group developed a novel optical system capable of observing thousands of galaxies at a time. This novel instrument was used with a large enough telescope, the Anglo-Australian 4 m telescope. In addition, Perlmutter advanced a new search strategy for supernovae. It consisted of collecting wide-field images just after the new moon, then subsequently collecting a second set of images just before the next new moon (and then subtracting the second set from the first set). The short time scale between the two sets of images ensured that the discovered supernovae would not have enough time to reach maximum brightness or start fading away. This gave time for large telescopes to record SN light curves. Over the years, the Berkeley group incorporated more astronomers and observatories from around the world to form a larger project known as the Supernova Cosmology Project (SCP), led by Perlmutter.

Almost simultaneously to the progress of the SCP project, a similar effort was developed. In 1989, motivated by the suggestion of Allan Sandage and by discussions at the UC Santa Cruz meeting on supernovae [158], Mario Hamuy, Nicholas B. Suntzeff, Mark Phillips, and Jose Maza joined efforts to begin a search for supernovae as a collaboration between the University of Chile and the Cerro Tololo Inter-American Observatory (CTIO) [159]. The survey, called the Calán/Tololo survey, aimed to rigorously test SN Ia as standard candles, using redshift as an accurate proxy for relative distance

The Calán/Tololo survey used the CTIO Curtis Schmidt telescope to take plates, each covering a field of 25 sq-deg on the sky. The plates were developed and dispatched the next morning to the University of Chile's Department of Astronomy, where they were examined. Objects that could be supernovae were observed the next day using a 0.9 m telescope equipped with a CCD camera, which had been exclusively set aside for that task. Between 1990 and 1993, 50 supernovae were identified, of which 32 were Type Ia [159].

In addition, the Calán/Tololo survey provided a uniform photometric and spectroscopic dataset of supernovae which led to the discovery of a method which used Type Ia supernovae as reliable standard candles: the "Phillips relationship" [160]. This method is based on the discovery of a tight correlation between the rate of decline from maximum light and peak luminosity of SN Ia light curves. Its application reduces dispersion in measurements of SN Ia peak luminosities.

In 1994, Brian P Schmidt, then a post-doctoral research associate at Harvard University, and Nick Suntzeff, a staff astronomer at CTIO, submitted a proposal to the CTIO called "A Pilot Project to Search for Distant Type Ia Supernova". This program (in Schmidt and Suntzeff's words) was "the next step in the Calán/Tololo SN survey". Old colleagues from the Calán/Tololo survey joined the program, as did astronomers from other parts of the world. The group, later named the High-Z Team (HZT), used the Keck telescopes in Hawaii and those of the European Southern Observatory (ESO) in Chile, in addition to having access to Hubble data. The HZT elected Brian P. Schmidt, then a postdoctoral fellow at Mt. Stromlo Observatory in Australia, to manage the team.

Under the leadership of Schmidt, the HZT developed a series of scripts to automatically subtract the massive amounts of imaging data. The procedure, in simple words, consisted of making the two images of two different epochs as identical as possible by aligning them, and then matching and scaling the image point spread functions between the two epochs. These two images were subtracted, and then the difference image was searched for new objects.

As a result of the HZT program, by 1998, the team had studied 48 SN Ia, of which 14 were high redshift. Similar results were achieved by the competing team, the SCP, by whom 42 high and 18 low redshift supernovae were studied. Measuring $q_0$, the deceleration parameter, had always been part of the aims of both the HZT and SCP programs. Both projects found in 1998 that high-redshift SN appeared to be 10% to 15% more distant than expected, which meant that the deceleration parameter $q_0$ was negative ($q_{1998} \sim$

–0.7518). In simple words, the universe's expansion was not slowing down, but rather accelerating, and there was a prevailing feeling that a cosmological constant was at work again in the field equations. The interpretation of this as a new acceleration was actually "decided" by vote at a scientific congress held in May 1998, the month during which Alan Riess of the HZT headed the first publication on the subject [161], and a short while later, a similar publication was authored by the SCP group [162]. In 2011, Saul Perlmutter, Brian Schmidt, and Adam Riess received the Nobel Prize in Physics for this discovery.

## 25. Mira Variables

In more recent times, other astrophysical *stars* have entered into the cosmological set. Mira variable stars have also been used as independent, intermediate-range distance (5–50 Mpc) indicators. Miras have absolute magnitudes comparable to those of Cepheids, and follow useful period–luminosity relations in the near infrared. As they are not in the optical range, measurements represent a challenge, as does the long luminosity periods of few hundred days. But Miras are ubiquitous and are present in all types of galaxy morphologies, so they are potentially interesting calibrators in galaxies where SN Ia are also present. Recently, Caroline D. Huang et al. [163] presented the first observations of Miras in an SN Ia host, yielding the first Mira-based calibration of the luminosity of SN Ia. Using this technique, they found an $H_0$ of around 73 ± 4 km/s/Mpc. The intention is to find Mira variables in a wide range of local SN Ia hosts in order to better determine the uncertainties.

## 26. The Deceleration Parameter $q_0$

In the second half of the twentieth century, the accepted models of the universe were based on general relativity under a pair of assumptions. The first was that the universe is homogeneous and isotropic on large scales, and the second assumption was that it contains normal matter, i.e., matter whose density falls directly in proportion to the volume of space which it occupies. Within this framework, in 1961, Allan Sandage, an astronomer close to Baade and Hubble, proposed observational tests to discriminate between selected world models (steady-state models, exploding cosmologies, etc.) [164]. These tests were often known as classical tests of cosmology.

Through the 20th century, the late universe was assumed to be dominated by a single component of matter with a density $\varrho_i$ compared to a critical density, $\varrho_{crit}$. The ratio of the average density of matter compared to the critical density is called the density parameter, $\Omega_M = \varrho_i/\varrho_{crit}$. This critical density is the value where the gravitational attraction of matter in the universe causes space to become geometrically flat. As an alternative, a universe could have open hyperbolic geometry if its density were below this critical value, and if it were above, the universe would be characterized by closed spherical geometry.

What Sandage's proposed tests tried to discern was the geometry of space. For this purpose, it was necessary to measure a deceleration parameter $q_0$ that was equivalently related to $\Omega_M$ through solutions to the Friedmann equation, assuming a universe consisting solely of normal matter. This relation turned out to be:

$$q_0 = \Omega_M/2$$

Astronomers needed observables to measure, and one of Sandage's tests included measuring the luminosity of an object as a function of its redshift $z$ [165].

$$D_L = \frac{c}{H_o q_o^2}\left[q_o z + (q_0 - 1)\left(\sqrt{1 + 2q_0 z} - 1\right)\right]$$

where $D_L$ is the luminosity distance, defined as the inverse square law of an object of luminosity, $L$, and observed flux, $f$:

$$D_L \equiv \sqrt{L/4\pi f}$$

These measurements in turn provided values for Hubble's constant and the average density of matter in the universe. Additional work by Sandage and others extended these types of measurements to further objects.

## 27. Gravitational Lenses

Until now, the techniques we have described to establish the $H_0$ value have been based on using the cosmic ladder to obtain a final estimate of absolute distance to the object being observed. Each rung of the ladder requires gauging a particular distance indicator, and each is subject to inaccuracies that contribute, in the end, to increasing the final uncertainty. Direct and more precise methods to estimate absolute distances, such as orbital parallax, are limited to nearby objects, and are difficult to apply to objects outside our own galaxy.

The ultimate solution would be finding primary indicators that are reliable and bright enough to directly calculate extragalactic distances, thus avoiding many rungs on the cosmic ladder. One of these techniques involves the use of gravitational lensing.

The use of lensing to measure the Hubble constant was first suggested by Sjur Refsdal as part of his PhD thesis. In 1964, Sjur Refsdal wrote a pair of fundamental papers, which were both communicated to the Royal Astronomical Society by Hermann Bondi. In the first, he described the properties of a point-mass gravitational lens, simplifying calculations previous made by Gavil A. Tikhov in 1937, arguing that geometrical optics could be used for gravitational lensing [166]. In the second paper, he studied the effect of a gravitational lens on a supernova lying far behind a distant galaxy, close to its line of sight. In this case, the gravitational lens was able to split the hypothetical supernova image into two images. The two different paths that light followed from the supernova reached the observer at different moments. Refsdal showed that this time delay, $\Delta t$, between both images can be used to determine the distance to the source $d_S$ and that to the lensing galaxy $d_L$. Refsdal showed that the delay is given by [167]:

$$\Delta t = k\left(\frac{d_L d_S}{d_S - d_L}\right).$$

The factor $k$ depends on the angular separation between images and the mass distribution of the lens. The angular separation is observable, while for the mass distribution, certain assumptions have to be taken. An additional correction to $\Delta t$ (not mentioned by Refsdal in that paper) must be applied due to the Shapiro time delay [168]. The Hubble parameter follows from the redshift ratio of the lens and the source. The difficulty of the method lies in that the mass distribution of the lensing galaxy or galaxies must be assumed to model the lens, and the angular separation of the lensed images might be difficult to measure.

In the same publication, Refsdal called attention to the potential importance of quasars in distance measurements using gravitational lenses. Here, it is pertinent to recall that some years before (1963) the first quasar, a "quasi-stellar", compact, very luminous, and distant source was identified by Maarten Schmidt [169].

Two more articles were published by Refsdal in 1968, where he suggested the use of lensing for testing cosmological theories such as the still-popular steady-state theory, and theories based on general relativity [170]. The second paper dealt with the conditions needed to determine the mass and distance of a star which acts as a gravitational lens, if the lens effect can be observed from the Earth and from at least one distant space observatory [171].

By that time, it had already been realized that light propagation in a real universe whose mass distribution is not homogeneous should differ from a homogeneous universe, because there are regions of space with greater mass density than others and, consequently, with different lensing effects. This fact must be considered. A third paper was published by Refsdal at the beginning of 1970 on the influence of lensing on the apparent luminosity of very distant light sources in static and flat universes with inhomogeneous mass distributions [172]. In this article, he studied the validity and limitations of this model, as well as possible extensions to expanding and curved universes.

In 1979, the first lensed quasar was discovered by Dennis Walsh, Robert F. Carswell, and Ray J. Weymann [173]. They reported the double quasar Q0957+561 as two quasar images with the same color, redshift, and spectra, separated by only 6.1 arc-seconds, produced by a gravitational lens. However, for a time, some radio astronomers disputed this interpretation, but further investigations confirmed it.

In this case (quasar Q0957+561), Refsdal's method could not be applied immediately due to two obstacles. The first was modeling the mass distribution of the lens, which turned out to be difficult as the lens consisted of at least two adjacent galaxies. The second difficulty was the measurement of $\Delta t$, which required long observation periods [174,175]. These were out of reach for many astronomers due to limited telescope allocation times and the usual problems of ground-based optical astronomy, such as the annual occultation by the Sun causing periodic gaps of 4 to 5 months in data acquisition. In addition, the flux variations in both images were small, and such measurements required extreme care and were difficult to carry out [176]. In addition to these difficulties, the works of Falco, Gorenstein, and Shapiro [177,178] and their coworkers [179] were key to realizing that the lens modeling was ambiguous and that some degeneracies existed in the parameter estimation. In particular, they pointed out that the Mass-Sheet Degeneracy (at that time known as Magnification Transformation) generates an ambivalence to estimate $H_0$ from the time delays between gravitationally lensed images. They also pointed out that this ambivalence can be addressed with a measurement of the lensing galaxy's velocity dispersion. The velocity dispersion constrains the lensing galaxy's mass, which breaks the degeneracy and leads to a point value for $H_0$ [180].

During 1991, Ramesh Narayan and other independent groups were the first to put Refsdal's idea into practice [181]. Narayan combined a model of mass distribution in the gravitational lens (Q0957+561) system, with measurements of the time delay, to obtain a value for $H_0$ of 37 ± 14 km/s/Mpc. Another independent estimate was made of the same quasar by Roberts et al. [182], and a value of 46 ± 14 (42 ± 14) km/s/Mpc was given for $H_0$. A third estimate was made by G. Rhee [183] that yielded 50 ± 17 km/s/Mpc.

All of these values had significant uncertainties. Part of the uncertainty was in modeling the mass distribution of the lensing system. The conspicuous, bright galaxy that apparently acted as the lens was part of a cluster of lensing galaxies. The time delay was also doubtful due to limited durations of the telescope observations.

In 1994, a team lead by Edwin Lewis Turner of Princeton gained access to the 3.5 m Apache Point Observatory in NM. From late 1994 through May 1995, every night, weather permitting, they observed the double quasar 0957+561 A B. They found the time delay of the trailing image B to be 415 days, with respect to the leading image A [184]. Their predicted time delay was confirmed in 1996, when the variability pattern between both images was 417 days. The result was published in 1997, together with their estimate of $H_0$ = 64 ± 13 km/s/Mpc. They argued that their estimate was of comparable quality to those based on more conventional techniques at that time. In the same paper, the authors pointed out the advantages of the gravitational lens method. Briefly, it is a geometrical method based on the well-understood physics of general relativity in the weak-field limit. It yields a direct, single-step method for measuring $H_0$ and thus avoids error propagation along the "distance ladder", which is no more secure than its weakest rung. It measures distances to cosmologically faraway distant objects, thus precluding the possibility of confusing a local with a global expansion rate. It gives an independent measurement of $H_0$ in two or more lensed systems with different sources and lens redshifts. This provides an internal consistency check of the obtained $H_0$ value, i.e., if a small number of time delay measurements all give the same $H_0$, this value can be regarded as correct with considerable confidence.

Further advances toward a model independent characterization of strong lenses have been developed in recent years [185], as have studies on the intrinsic degeneracies of the gravitational lensing formalism [186]. With these works, one was finally able to derive the most general class of lensing degeneracies, make physical sense out of the mathematical parameters, and put them into the overall context of related cosmological work. These results gave a more definitive basis on which to understand strong lensing in cosmology. As Jenny Wagner put it, "…after this major breakthrough, the constraints on $H_0$ from lensing that had been accused of being a black-box, can now be put on solid grounds" [187].

The gravitational lens method increased in importance as more gravitationally lensed quasars were discovered for which the mass distribution of the lens was simpler than that of Q0957+561, but that was to occur in the twenty-first century.

## 28. Megamasers

Another distance indicator that is independent of the cosmic ladder is the megamaser technique, i.e., a maser powerful enough for us to see at cosmic distances. This powerful emitter turned out to be a mega $H_2O$ maser produced inside vast clouds of water molecules located in the vicinities of galaxy centers and pumped by infrared radiation. These masers, as we shall see, facilitated a more precise distance assessment of the NGC4258 galaxy.

In 1995, a group of American and Japanese astronomers found compelling evidence for the existence of a supermassive compact object at the center of the NGC4258 galaxy [188]. This evidence was gathered using the very long baseline array (VLBA). The VLBA is a is a system of ten 25-meter radio telescopes distributed along the USA's territory and operated remotely from Socorro, New Mexico. This set of antennas works together as a very extended baseline interferometer. Its baseline is about 8600 km, thus producing superb angular resolution.

The team measured the intensity of the narrow spectral line emissions (1.35 cm) of water maser sources. During this surveillance, they registered spikes of $H_2O$ maser emissions, with each spike corresponding to a lump of $H_2O$ masering material. These spikes came from masers with radial velocities equal to the average velocity of the galaxy, but they also found spikes from masers with higher and lower radial velocities. The simplest explanation is that there is a disk of $H_2O$ gas orbiting a central mass in NGC4258. The high and low velocity spikes originated from masers at the edges of the disk. From the dynamical characteristics of the Keplerian motion of the disk, the group deduced that there was a black hole (BH) at the center of rotation, with a mass of at least $3.6 \times 10^7$ $M_\odot$. The discovery made the news. Incidentally, at nearly that time, there were independent investigations by Reinhard Genzel and Andrea Ghez regarding the existence of a BH at the center of our Milky Way that, years later, resulted in Nobel prizes for this latter pair of scientists.

It was not until 1999 that further investigations conducted by members of the same original team of astronomers introduced a novel geometrical method to infer the extragalactic distance from their direct measurement of orbital motions of $H_2O$ masers in the disk of masing clouds surrounding the nucleus of the NGC4258 (M106) galaxy [189].

This time, astronomers reported that the disc had an estimated a radius of 0.13 pc, and from its angular radius (4.1 mas), they deduced its distance: $6.4 \pm 0.9$ Mpc. The novelty of this geometric method was that required no assumptions or calibrations, just precise measurements of angles and velocities. The method turned out to be much more accurate than other previous distance determinations of NGC4258.

In subsequent observations of the mega $H_2O$ masers in NGC 4258, measurements continued to be refined over the years. Table 3 shows the values of the distance to the NGC 4258 galaxy, with uncertainty estimates. The reader will notice that, currently, the margin of error is small. This, as we shall later comment, has also resulted in a dramatic reduction in uncertainty surrounding the Hubble constant.

Over the 25 years after the first report of the distance to NGC 4258 in 1995, the number of VLBI observations and geometric distance estimates increased. These new estimates were based on modeling the Keplerian orbits of $H_2O$ masers orbiting around the BH using larger data sets. Table 3 shows these newer estimates of distance to NGC 4258, with the last three reports showing only very small changes in the estimated distance, but with successive improvements in the uncertainty.

**Table 3.** Estimates of the distance to NGC 4258.

| Reference | Distance (Mpc) |
|---|---|
| Miyoshi et al. (1995) [188] | 6.4 ± 0.9 |
| Herrnstein et al. (1999) [189] | 7.2 ± 0.5 |
| Humphreys et al. (2013) [190] | 7.596 ± 0.228 |
| Riess et al. (2016) [191] | 7.540 ± 0.197 |
| Reid et al. (2019) [192] | 7.576 ± 0.112 |

Also using this technique, the Megamaser project (MCP) [193] aims to determine the Hubble constant. They have been working for over a decade, and recently, they found a value of $H_0$ = 73.9 ± 3.0 km/s/Mpc [194] based on measurements of five megamaser galaxies in the Hubble flow. Their measurements are consistent with the high end of $H_0$.

## 29. The Sunyaev–Zeldovich Effect

Another way to estimate the Hubble constant is through looking at distortions in the CMB radiation caused by the scatter of hot gas in galaxy clusters. The CMB light travels from the last scattering surface to us, and in between, it finds a series of clusters that possess a hot gas of electrons that, in turn, scatters the CMB light, provoking a CMB brightness decrement in the direction of the cluster. This effect was proposed in 1970 by R. A. Sunyaev and Ya. B. Zel'dovich [195,196], and it is named to honor these two scientists—the SZ effect, for short. There are two effects: the thermal, which is caused by energetic electrons in the cluster, and the kinematic, caused by the relative motion of the cluster with respect to the Hubble flow.

In the late 1970s, the idea was put forward that a comparison of the thermal SZ effect with the X-ray properties of a cluster, also caused by high-energy electrons, could help to constrain cosmological parameters such as the Hubble constant, among others. These two effects have different degeneracies of their amplitudes, so its comparison could help to break parameter dependencies. The method is also independent of a standard candle or ruler, relying only on the properties of highly ionized plasma, with temperatures around 10 keV. Using the same density of electrons in the cluster, one can obtain the angular diameter distance, and hence infer the Hubble constant. For a summary of the physics of this technique and related works up to the early 2000s see the website cited in [197].

The first works to consider this idea were Cavaliere, Danese, and de Zotti [198] in 1977; Cowie and Perrenod in 1978 [199]; Gunn et al. in 1979 [200]; Silk and White in 1978 [201]; and Cavaliere et al. in 1979 [202]. Early estimations were made by Birkinshaw in 1979 [203]. Later, in the 1990s, M. Birkinshaw and J.P. Hughes [204] measured it in a finer way, though the uncertainties at that time were still big (±7 km/s/Mpc). A decade later, other measurements were carried out, for example by Reese et al. in 2000 [205] and 2002 [206]; Patel et al. in 2000 [207]; Mason et al. in 2001 [208]; Sereno in 2003 [209]; Udomprasert et al. in 2004 [210]; Schmidt et al. in 2004 [211]; and Jones et al., who used ground-based radio interferometers, in 2005 [212]. In that epoch, the value of the Hubble constant was still around 50–100 km/s/Mpc, not only because of the uncertainties in the measurements, but also because of the large spread of the central values obtained via different methods.

The SZ technique has been further refined with the advent of CMB probes such as WMAP and, later, SPT (South Pole Telescope); ACT (Atacama Cosmology Project); and Planck. These are presently obtaining much smaller uncertainty values (±3 km/s/Mpc) [213].

## 30. $H_0$ Measurements with CMB Probes and BAO

There are two other modern techniques used to measure the Hubble constant, both of them related to the physics of the early universe's acoustic oscillations: CMB and BAO. CMB measurements are now an important basis of modern cosmology. The CMB is the fingerprint of the universe, providing us information about not only the time it formed, but also the initial conditions of the universe long before the first second of life. As if that were not enough, since this light comes from such remote distances and from moments so far away, on its journey to us, it collects information about the properties of the universe on its path. The general features of this very important fingerprint were theoretically calculated prior to its observation using the Big Bang model. The light from the CMB comes from the farthest region and, therefore, from the earliest time in our universe that we will ever be able to directly see, no matter how powerful future telescopes or satellites become. This is because all the light that had previously interacted with the electrons and protons in a very efficient way has ceased to exist, although new light is generated. As the universe cooled, the temperature dropped and the light stopped interacting; the free electrons combined with protons to form neutral hydrogen atoms (one proton and one electron). When this happened, the universe was about 380,000 years old. The last scattered photons formed the so-called last scattering surface, located 13,400 million of years in our past. The light stemming from it traveled to us in an almost unperturbed way. The CMB light was measured in 1964 by Arno Penzias and Robert Wilson, who deserved the 1978 Nobel Prize in Physics. They observed a noise in their microwave antenna, and after examining possible sources of noise, even down to the pigeon droppings on the antenna, they deduced that it was an observed noise coming from a celestial source of extragalactic origin. The correct interpretation of cosmic background radiation was made by scientists in the Robert H. Dicke group, which included P. J. E. Peebles, P. G. Roll, and D.T. Wilkinson, who had been seeking to measure it at that time, in addition to the group of Russian scientists from the Yaakov B. Zel'dovich group from Moscow, which included A. G. Doroshkevich and I. D. Novikov. The measured temperature was approximately −270 °C, about 3K above absolute zero. Time passed, and towards the end of the 1980s, a modern version of Penzias and Wilson's experiment was carried out. This experiment was commanded by one of the authors of this article (G.F.S.) and his collaborators, and was carried out with the COsmic Background Explorer (COBE) satellite. Thus, this began a new experimental era of high precision in cosmology. The COBE team, in the early 1990s, revealed with great accuracy that the universe is homogeneous and isotropic, but not entirely. Tiny anisotropies discovered by COBE, one part in a hundred thousand, were responsible for the subsequent formation of stars, galaxies, galaxy clusters, and all the large-scale structures of our universe. Furthermore, COBE found a blackbody spectrum from the CMB indicating that the early universe was in thermodynamic equilibrium, with (almost) the same temperature everywhere and in a homogeneous state. These were two of the most important discoveries in 20th-century astronomy, and they are why George F. Smoot and John Mather were awarded the 2006 Nobel Prize in Physics.

The origin of the minuscule anisotropies was most likely due to the primordial fluctuations of a quantum field which was present in the very early universe, during the inflationary era. Modern quantum field theories, along with cosmological models, help to understand how small fluctuations evolved at cosmic scales, and then how COBE is able to detect them tens of billions of years later. This early period of inflation was a stage of accelerated growth of our universe, and not only gave rise to the formation of primordial fluctuations, but also, although it seems contradictory, explains why our observable universe is so homogeneous and isotropic. Inflation was postulated in the early 1980s independently by Alex Starobinsky and Alan Guth, and was further developed by Andrei Linde and many others. It is an essential part of the Big Bang model in the first instances of the existence of the universe.

The COBE satellite and other cosmological probes, such as the BOOMERANG and MAXIMA hot air balloons, launched in Antarctica in the late 1990s; the more recent WMAP (Wilkinson Microwave Anisotropy Probe) satellite launched in 2001; and the Planck satellite in 2012, have confirmed these peculiar characteristics of the origin of our universe. The CMB not only provides information on the fundamental fluctuations, but also on the behavior, of the universe at that time. Regions of the space that subtend an angle of less than one degree from us to the last scattering surface have undergone oscillations in their growth due to the interactions between protons, electrons, and light just before the hydrogen atoms were formed. The competition between the force of gravity, which attempted to bring all the particles together, and the pressure force of light, which fought against it, generated these oscillations to the primordial plasma, imprinting a unique fingerprint on light and matter. At present, when we measure the acoustic pattern of the CMB light, we are able to measure different cosmological parameters that are fitted to CMB maps in a delicate manner, given the precision of Planck. An efficient way to compare maps with model parameters is through the use of the Monte Carlo Markov Chain (MCMC) technique for the model predictions. The theoretical prediction is performed by solving the Friedmann equation for the background evolution together with perturbed Boltzmann equations for each component of the universe: "baryons" (nuclei and electrons), light, neutrinos, dark matter, and dark energy (or the constant $\Lambda$). There are different Boltzmann solvers, starting from the pioneering code of Uros Seljak and Matias Zaldarriaga, called CMB-Fast, in the mid 1990s, as well as many others that came later. Today, the most used ones are CAMB and CLASS, which are coupled to MCMC solvers. There are six cosmological parameters to solve (the present amount of baryons and dark matter; the angle of the CMB acoustic scale; the reionization parameter; and two inflation signatures: the amplitude of primordial perturbations and the spectral index). This may depend on additional parameters for extended models, depending on, e.g., the universe's curvature, the number of relativistic species, the mass of neutrinos, etc. The Hubble constant is tightly constrained in combination with density parameters, so this technique permits us to indirectly determine it. As already mentioned, the preferred value for the flat ΛCDM model is $H_0 = 67.4 \pm 0.5$ km/s/Mpc [214].

Once light decouples from plasma, baryonic matter begins to fall to potential wells of DM, attracting DM to it while keeping the oscillatory patterns formed by the plasma. This imprints on the matter clustering a specific feature at a large scale that serves as a standard ruler and tells us about the size of the horizon at last scattering (or drag time, to be more specific). These baryon acoustic oscillations (BAOs) represent an amazing pattern of cosmological dimension in which over-densities in the matter field are found, as a rule, at a comoving size every ca. 150 megaparsecs. These acoustic oscillations were predicted long before they were measured. BAOs were estimated in the mid-1990s, and their measurement was reported in 2005 for first time by two independent groups, the Sloan Digital Sky Survey (SDSS) collaboration led by Daniel Eisenstein [215] and the 2dF collaboration led by Shaun Cole [216]. This was another amazing and very beautiful success in science. By localizing galaxies in the 3D universe through the BAO technique, one is able to compute the distances to the BAO feature at different times in the past. In the 2000s, the 2dF, 6dF, WiggleZ, and SDSS (BOSS) collaborations measured the BAO feature and computed the distance to it at different redshifts. Again, the Hubble constant was indirectly measured, given a (ΛCDM) model. The most refined measurements to date were obtained from the eBOSS collaboration, in which different tracers were used (see, e.g., [217]), and their results were in accordance with those of Planck.

## 31. Gravity Waves Standard Sirens

Astrophysical events at cosmological distances may also be useful in determining the Hubble constant. We will see examples of these in the following sections.

A new, interesting, and different method to determine $H_0$ is through gravity waves emitted from the coalescence of neutron stars at the final stages of binary systems. In principle, coalescence of black holes emits gravity waves as well, but does not provide redshifts nor precise distances. An electromagnetic counterpart is needed, and that is why one employs neutron stars. This method is considered to be applicable to every neutron coalescence. The name "standard siren" was picked to distinguish it from the known standard candles [218]. This method was proposed long ago by Bernard Schutz [219], but its realization materialized just few years ago. The method relies on the physics behind the collapse of the binary, and it is free of the systematics of local distance scales.

This technique provides luminosity distances, and because they are well modeled during its coalescence phase, their uncertainties could be potentially small. On 17 August 2017, LIGO/Virgo collaboration detected a pulse of gravitational waves, called GW170817, associated with the merging of two neutron stars in NGC 4993, an elliptical galaxy in the constellation Hydra located 43 Mpc from us. The collaboration paper [220] reported a Hubble constant of $70^{+12}_{-8}$ km/s/Mpc. With this, we are entering into a new era of gravitational wave multi-messenger astronomy. The associated uncertainties will be greatly reduced in the coming years as more events of this type are measured.

One can also make a pure gravitational wave estimation of Hubble's constant using neutron star–black hole mergers [221]. $H_0$ can be derived purely from the gravitational waves of neutron star–black hole mergers. This new method provides an estimate of $H_0$ spanning the redshift range of $z < 0.25$, with the sensitivity of current gravity waves and without the need for any afterglow detection. The authors employed the inherently tight neutron star mass function together with the merge's waveform amplitude and frequency to estimate the distance and redshift, respectively, thereby obtaining $H_0$ statistically. Their first estimate is $H_0 = 86^{+55}_{-46}$ km/s/Mpc for the secure neutron star–black hole events GW190426 and GW200115. One expects that with ten more such events, one may reach a precision of $\delta H_0/H_0 \lesssim 20\%$.

## 32. Black Hole Shadows

Recently, another alternative method to determine the cosmic distance scale has been proposed to measure the Hubble constant [222]. It is based on the measurement of the shadows of supermassive black holes in order to use them as standard rulers; a historical account of black hole shadows can be found in this reference [223]. With a known mass of a black hole, its shadow is unique within GR and has a known physical size. By measuring its angular size, its angular diameter distance can be determined, which in turn depends on $H_0$. Technically, however, the problem is intricate, since light paths must be computed under the influence of the local action of gravity and, at the same time, in an expanding background. This technique is applicable to the late universe for small redshifts ($z \leq 0.1$), for large black hole masses ($\geq 10^9$ solar masses), and potentially also for much larger redshifts in the coming years. Even though the angular diameter distance decreases for higher redshifts, the angular size of a shadow is expected to increase for high redshifts [224]. This scope will allow us to study angular diameter distances as functions of time. The current precision of H0 is on the order of 10% according to data such as those from the Einstein Horizon Telescope [225], or a small (few) percentage when considering probes in the near future [226–228]. A note of caution has been raised on this technique regarding the difficulty of obtaining reliable estimates of a black hole's mass; achieving the required angular resolution; and having sufficient knowledge of high redshift accretion dynamics, especially challenging high redshift measurements [229]. Other possible effects may come from alternative models to $\Lambda$CDM in which pressure singularities may exist, which change the shadows of black holes and, therefore $H_0$. This could eventually resolve the Hubble tension [230,231].

## 33. Fast Radio Bursts

Fast radio bursts (FRBs) are millisecond-duration pulses of radio emission observed at frequencies from hundreds of MHz to a few GHz, now recognized to have a cosmological origin. Their physical source is, however, still under debate, and there have been many progenitor models proposed. FRBs have the potential to act as cosmological probes, and in particular to determine the Hubble constant at redshifts $z < 1$, among other applications. FRB pulses scatter as they travel through the ionized intergalactic medium, with the inferred total dispersion measurement being a powerful probe of the column density of ionized electrons along the line of sight. The dispersion measure and the signal-to-noise ratio of the pulse, together with the redshift associated with the host galaxy of the burst, provide a direct constraint on $H_0$. One approach to computing $H_0$ is to assume that the FRB energetics do not depend on redshift, being a kind of standard candle, and the sensitivity to $H_0$ is given through the signal-to-noise ratio of the pulse. It appears unlikely that they are standard candles unless there is a correction factor, since repeater pulses are far from identical from one pulse to the next. The other approach considers the cosmic contribution to the FRB dispersion measure, whose average depends directly on both the Hubble constant and the baryon density parameter; thus, synergies with CMB and nucleosynthesis results could help to constrain $H_0$. In the study of C.W. James et al. [232], a detailed methodology and bias estimates are provided. Using a sample of 9 FRBs, Steffen Hagstotz et al. [233] found a Hubble constant of 62.3 ± 9.1 kms/Mpc/s, whereas Qin Wu et al. [234] used 18 localized FRBs and found an $H_0$ with a smaller uncertainty of $68.81^{+4.99}_{-4.33}$ kms/Mpc/s. Finally, using a sample of 16 localized and 60 unlocalized FRBs, C.W. James et al. [235] found a best fit of $73^{+12}_{-8}$ kms/Mpc/s. This new approach is certainly developing, and will surely provide more stringent constraints on $H_0$ in the near future from the CHIME catalogs [236], among others.

## 34. The Current Situation and Final Remarks

By the early 1990s, many estimates of the Hubble constant had already been made using several methods. As a result of these measurements, a dichotomy of results surged. Values for $H_0$ were grouped into one of two separate intervals. In the first set of the two, dubbed the "long" timescale, the value of $H_0$ was situated in the interval between 40 to 60 km/s/Mpc, while in the second set, labeled the "short" timescale, $H_0$ was located between 80 to 100 km/s/Mpc (i.e., a large value of the Hubble constant and a "short" timescale for the cosmological expansion). This indicated a "factor-of-two" difference!

Recognized astronomers such as Allan Sandage, and Gustav A. Tammann, and some supernova theorists have argued that the value of the Hubble constant is in the range of the long timescale, while Gérard de Vaucouleurs and others have championed for the short-scale value. Debates between defenders of both positions lasted several years and were inconclusive.

On the one hand, short-timescale critics brought up the "Malmquist bias" as one of the most important sources of their opponents' miscalculations. This bias, in a brightness-limited survey of luminous objects in the sky, is an unintentional type of censorship where luminous objects below a certain apparent observational brightness threshold are excluded. They cannot be observed. This unintentional exclusion is the source of the bias. On the other hand, "long"-scale critics have mentioned accumulative systematic errors made by their opponents in the processes of measurements and treatment of data originally used to gauge the ladder scale.

Over the years, this controversy created an unsatisfactory and uncomfortable situation. Astronomers could not reach a consensus on how they could adjust their data to account for various effects that might lead to bias in their observations. Differences in the choice of secondary distance indicator methods were at the root of almost all the current debate at that time concerning the value of the Hubble constant.

Later on, with the advent of the HST, uncertainties diminished. Using different techniques, Wendy Freedman et al. reported the final results from the measured a value of $H_0$ of around 72 to ca. 10% accuracy [237]. This value was in the middle of the short and long timescales.

Alternatives arose to end this stagnation. They were proposed by independent groups, and were not based on climbing the rungs of the cosmic ladder. They included, namely, gravitational lensing, the S-Z method, and CMB and BAO analysis.

In recent years, Cepheids techniques have made progress in different directions in different projects; for an account of this, see reference [143]. It turned out that methods anchored to distance ladder techniques (Cepheids, supernovae, Mira variables) have been finding Hubble constant values in the range of 72–74 km/s/Mpc using the standard flat ΛCDM model. Recent works of the late-universe teams, such as the SH0ES collaboration, which used the distances to supernovae calibrated with Cepheids, reported an $H_0$ of ~73 ± 1 km/s/Mpc; these results were confirmed using the latest James Webb Space Telescope (JWST) measurements by Adam Riess et al. [238]. Other methods, such as strong lensing techniques, also obtain a Hubble constant in this interval. For instance, the H0LICOW collaboration reported $H_0 = 73.3^{+1.7}_{-1.8}$ km/s/Mpc at 2.4% precision using the light from six multiply-imaged quasar systems [239]; although different results have been obtained since the modeling of the lens, mass distribution is an important systemat-ic. These methods use late-time physics. On the contrary, early-universe physics at last scattering are anchored to the sound horizon, yielding CMB and BAO results that indicate smaller values of $H_0$, around 67 km/s/Mpc. The uncertainties reported in these papers show a discrepancy of around four sigma, or larger in some cases. In Figure 12, we show the curves of the ΛCDM model that predict SH0ES data together with clustering data from the BOSS collaboration, which measured the Hubble expansion rate as a function of redshift. The differences are clear.

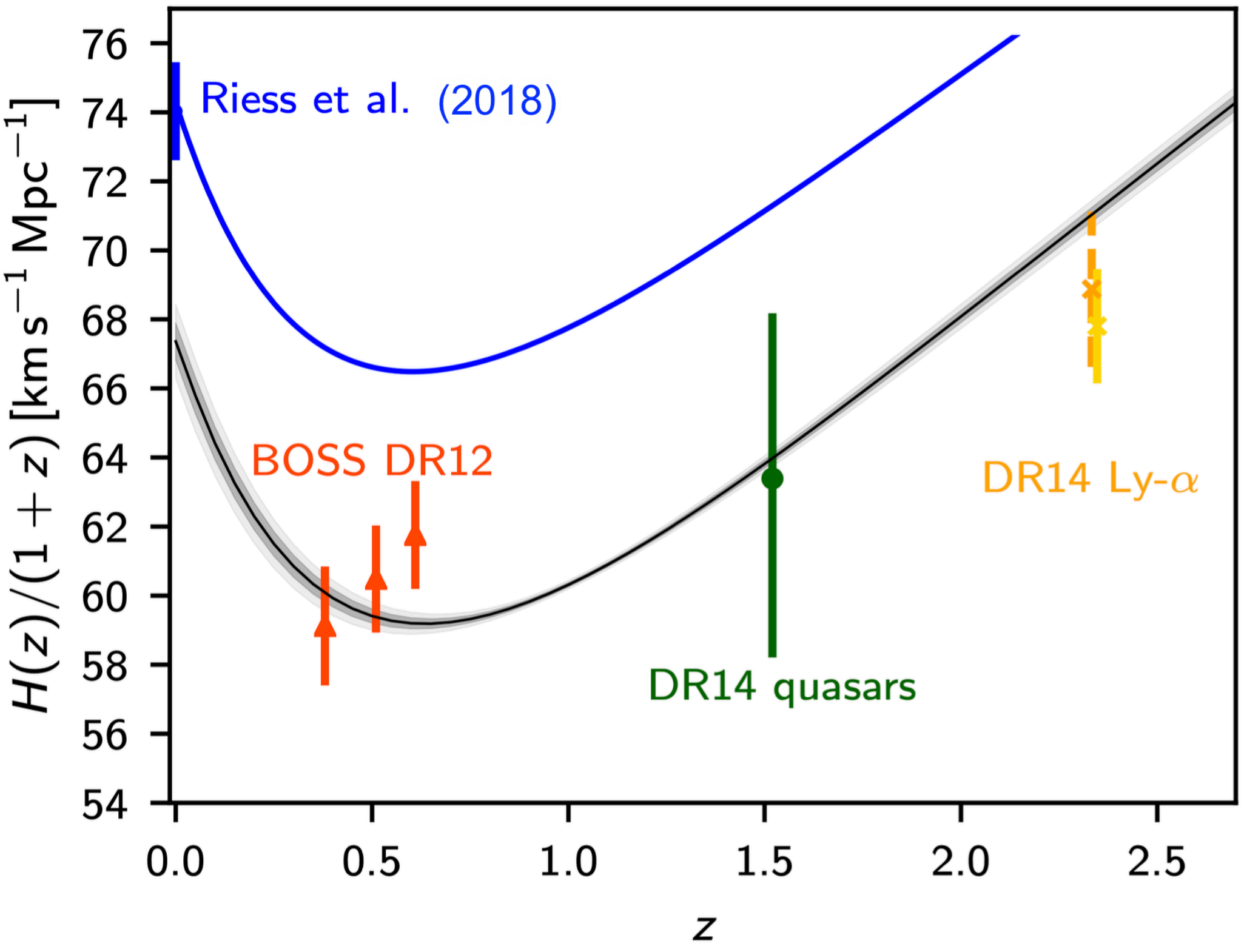

**Figure 12. Hubble expansion rate,** as observed in the present (2023) and earlier (vs. redshift), along with the best-fitted cosmological model predictions from CMB observations (the lower curve). The upper curve is the model used for fitting the "nearby" supernova data, which extended out to z ~1. The actual data fit fairly tightly on this curve. Note that by quoting as the Hubble Constant $H_0$ one suppresses the model dependence of the nearby observations. Doing so allows us to see that the model would need to have "new physics" to bring these into agreement. If we assume some form of continuity in the expansion of the universe, then "new physics" has to jump between the light-based standard candles and the size-based standard candles (cosmic rulers). Credit: Figure was modified by us after a plot from reference [214]. The blue point corresponds to the measurement in in ref. [240].

On the other hand, based on TRGB the CCHP collaboration recent results by Freedman et al. [241], lies somewhat in the middle of the tension, with an $H_0$ around 70 ± 2, as Figure 13 shows. These results ameliorate the discrepancy, and eventually may point to a solution within the ΛCDM model. A likely way to resolve the apparent discrepancy is for at least two of the methods to have additional systematic errors that are underestimated. The big question is which are those two or three out of three?

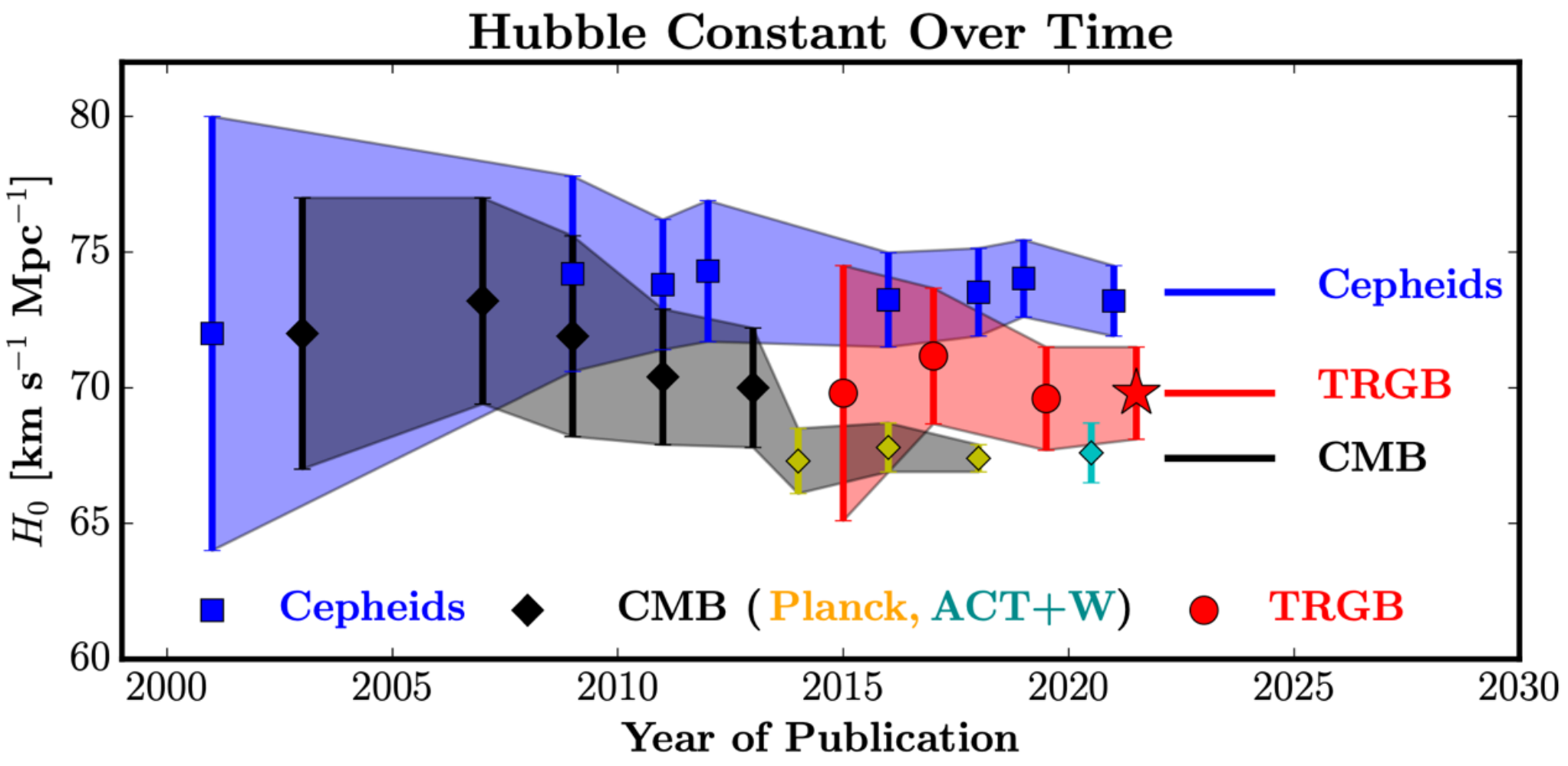

**Figure 13.** Determination of Hubble constant using different techniques in the last two decades, with shadows showing the associated uncertainties. Credit: Wendy Freedman, reference [241].

In the meantime, this controversy has inspired many theoretical papers to attempt to resolve this discrepancy. Recent papers have appeared to explain the discrepancy of the Hubble constant determination using late- and early-universe physics. They are based on different types of physics, anchored in different scales, but both are apparently correct. A compilation of recent $H_0$ measurements is shown in Figure 14. It is clear that early physics determinations were smaller than those of late physics, except for those measurements from the TRGB obtained by CCHP. See further details in reference [242], and this reference [243] added during galley proofing corrections that accounts for the different modern techniques to measure $H_0$.

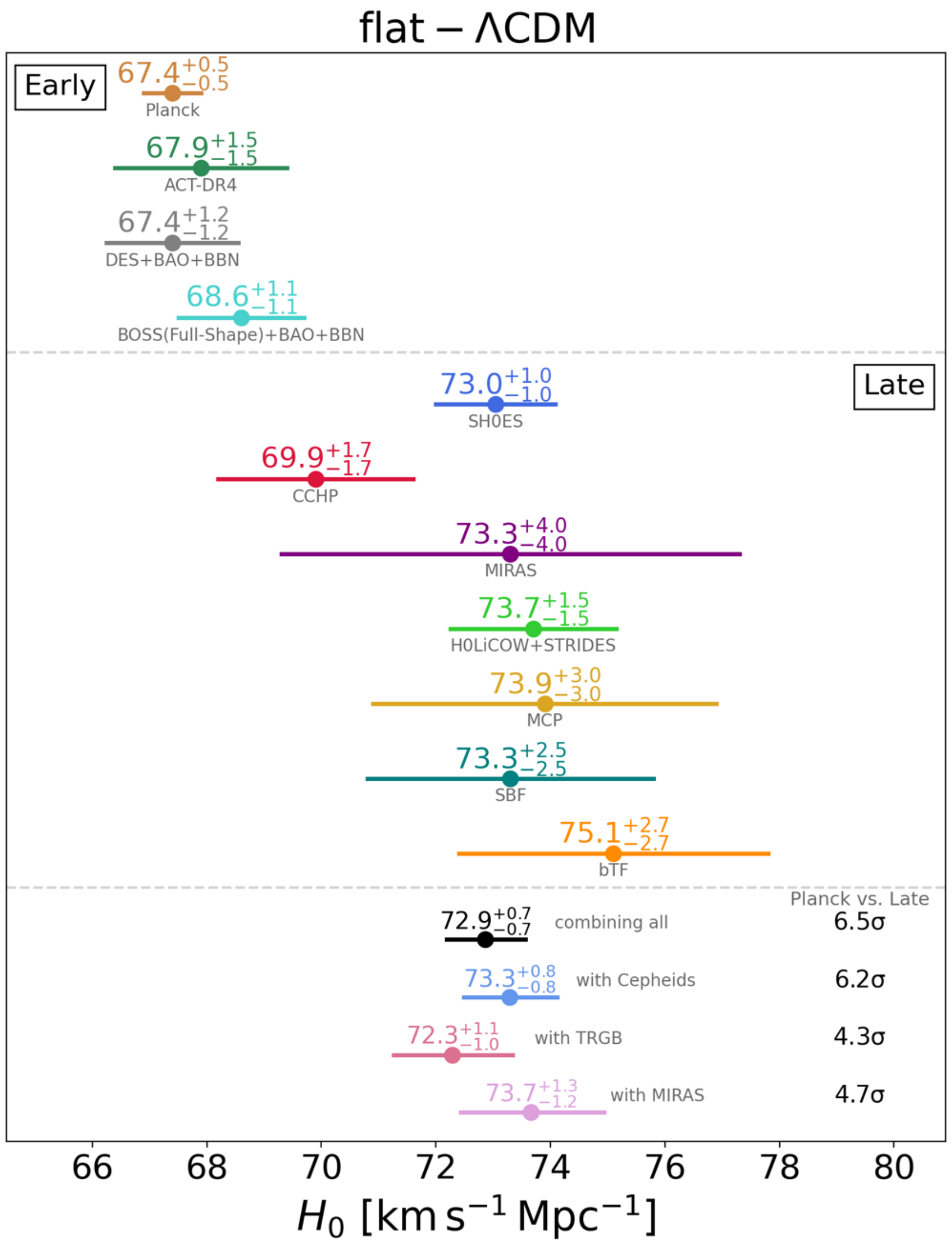

**Figure 14.** The spread of the Hubble constant measurements performed in recent years, showing small values for CMB and clustering physics (**upper** panel) and large values for different rungs used in late measurements (**middle** panel). Also, different combinations of methods are shown, as well as their discrepancies with the early-physics results (**bottom** panel). Credit: Vivien Bonvin and Martin Millon [244].

In the near future, one hope lies in the emerging data that the JWST is certainly collecting. The JWST is making new infrared and more precise optical observations available. As stated by Freedman and Madore [143], the JWST has a resolution three times that of the HST, with nearly ten times the sensitivity. Thus, it will reduce the Cepheids, TRGB, and Miras systematics, among others, in order to better determine the Hubble constant. At the same time, new CMB experiments are nearly ready, such as The Simons Observatory and plans such as CMB-S4. In addition, new BAO probes such as DESI, EUCLID, and LSST are underway; these will improve our precision in calculating the Hubble constant. Will these coming measurements relax the Hubble tension or increase it? A more precise answer on the expansion rate of the universe and the Hubble constant will make it clear whether the ΛCDM model will persist or whether new physics will need to be added to it, and new theoretical horizons will surge. In fact, there are already plenty of possible explanations to the Hubble tension; see, for instance, a review of solutions [245,246]. Also, several hints have been put forward [247] suggesting that modifications to both the early and late universe are necessary in order to solve the problem of Hubble tension.

In conclusion, we will remark that there is a prejudice introduced when one labels a parameter that changes with time as a constant. In fact, the Hubble constant is the value the parameter H(z) has at the present time, but it will have a very different value if measured 10 billion years before or after the present time. $H_0$ is not a fundamental physical constant. It provides a scale for the present-day universe, as it is a reference. There is even some evidence that $H_0$ might have a decreasing trend when computed using data at higher redshifts (see reference [239] and others [248, 249]). Now, its different estimations resulting from diverse standards based on different anchors may be indicative of some new physics, although this also might simply be a difference due to different systematic errors in the various techniques. We believe that time, as well as additional and improved observations, will settle the argument, just as occurred with the Great Shapley–Curtis debate.

**Author Contributions:** All authors contributed in an equal manner. All authors have read and agreed to the published version of the manuscript.

**Funding:** J.L.C.-C. was funded by CONAHCYT grant number 283151.

**Data Availability Statement:** No new data were created or analyzed in this study. Data sharing is not applicable to this article.

**Acknowledgments:** This research made use of NASA's Astrophysics Data System Bibliographic Services.

**Conflicts of Interest:** The authors declare no conflicts of interest.

## Notes

1. For thousands of years people had referred to the "fixed stars". In astronomy, the fixed stars (Latin: *stellae fixae*) are the luminary points, mainly stars, that appear not to move relative to one another against the darkness of the night sky in the background. This is in contrast to those lights visible to naked eye, namely planets and comets, that appear to move slowly among those "fixed" stars. So, we had thousands of years prejudice for a static Universe and a human lifetime natural time scale while the stars move on time scales of millions of years or more.
2. What Einstein did not imagine is that the cosmological constant that he would later regret having introduced, and called his "biggest blunder" would later represent the dark energy that is an essential ingredient of the modern $\Lambda$CDM model.
3. All initial cosmological models assumed spherical (actually homogeneous and isotropic) symmetry in order to make cosmology a tractable issue. In the first half century this was an assumption but later radio surveys and most importantly the isotropy of the Cosmic Background Radiation have justified and improved this assumption. The CMB limits are uniformity to the part in 100,000 level or better. This is sufficient to find the first order solutions and treat the rest with perturbation theory.